\documentclass[journal]{IEEEtran}[12pt]
\ifCLASSINFOpdf
\usepackage[pdftex]{graphicx}
\graphicspath{{../pdf/}{../jpeg/}}
\DeclareGraphicsExtensions{.pdf,.jpeg,.png}
\else
\usepackage[dvips]{graphicx}
\graphicspath{{../eps/}}
\DeclareGraphicsExtensions{.eps}
\fi\usepackage{graphicx}

\usepackage{graphics}
\usepackage{epsfig}
\usepackage{epstopdf}
\usepackage{stfloats}
\usepackage[cmex10]{amsmath}
\usepackage{algorithmic}
\usepackage{array}
\usepackage{mdwmath}
\usepackage{mdwtab}
\usepackage{graphicx}
\usepackage{subfigure}
\usepackage{color}
\usepackage{amsfonts,amssymb}
\usepackage{hyperref} %文献和公式自动跳转！！
\usepackage{multirow}
\usepackage{diagbox}
\usepackage{caption}
\usepackage{bbding}
\usepackage{amsfonts,amsthm,array} %remark 加黑
\usepackage[ruled]{algorithm2e}
\usepackage{makecell}%表格内换行
\usepackage{lineno} %显示行号

\usepackage{bbm}
\usepackage{amsmath}%对齐

\begin{document}
	%\linenumbers
\title{On Secure mmWave RSMA Systems\thanks{Manuscript received.}}

\author{Hongjiang~Lei, %~\IEEEmembership{Senior Member,~IEEE,}
	Sha~Zhou,
	{Xinhu~Chen,}
	Imran~Shafique~Ansari, \\%~\IEEEmembership{Member,~IEEE,}
	Yun~Li,
	Gaofeng~Pan, 
	and~Mohamed-Slim~Alouini %,~\IEEEmembership{Fellow,~IEEE}
\thanks{This work was supported by the National Natural Science Foundation of China under Grant 61971080. (Corresponding author: \textit{Hongjiang Lei}.)}
\thanks{Hongjiang~Lei is with the School of Communication and Information Engineering, Chongqing University of Posts and Telecommunications, Chongqing 400065, China, also with Chongqing Key Lab of Mobile Communications Technology, Chongqing 400065, China (e-mail: leihj@cqupt.edu.cn).}
\thanks{Sha~Zhou and Xinhu~Chen are with the School of Communication and Information Engineering, Chongqing University of Posts and Telecommunications, Chongqing 400065, China (e-mail: cquptzhous@163.com;{chenxhcqupt@163.com}).}
\thanks{Imran~Shafique~Ansari is with James Watt School of Engineering, University of Glasgow, Glasgow G12 8QQ, United Kingdom (e-mail: imran.ansari@glasgow.ac.uk).}
\thanks{Yun~Li is with Chongqing Key Lab of Mobile Communications Technology, Chongqing 400065, China (e-mail: liyun@cqupt.edu.cn).}
\thanks{Gaofeng~Pan is with the School of Cyberspace Science and Technology, Beijing Institute of Technology, Beijing 100081, China (e-mail: gaofeng.pan.cn@ieee.org).}
\thanks{Mohamed-Slim~Alouini is with CEMSE Division, King Abdullah University of Science and Technology (KAUST), Thuwal 23955-6900, Saudi Arabia (e-mail: slim.alouini@kaust.edu.sa).}
}

\maketitle
%%%%%%%%%%%%%%%%%%%%%%%%%%%%%%%%%%%%%%%%%%%%%%%%%%%%%%%%%%%%%%%%%%%
\begin{abstract}
	
Millimeter-wave (mmWave) communication is one of the effective technologies for the next generation of wireless communications due to the enormous amount of available spectrum resources.
Rate splitting multiple access (RSMA) is a powerful multiple access, interference management, and multi-user strategy for designing future wireless networks.
In this work, a multiple-input-single-output mmWave RSMA system is considered wherein a base station serves two users in the presence of a passive eavesdropper.
Different eavesdropping scenarios are considered corresponding to the overlapped resolvable paths between the main and the wiretap channels under the considered transmission schemes. 
The analytical expressions for the secrecy outage probability (SOP) are derived respectively through the Gaussian–Chebyshev quadrature method.
Monte Carlo simulation results are presented to validate the correctness of the derived analytical expressions and demonstrate the effects of system parameters on the SOP of the considered mmWave RSMA systems.

\end{abstract}

\begin{IEEEkeywords}
	Millimeter-wave,
    rate splitting multiple access,
    uniform linear array, 
	secrecy outage probability.
\end{IEEEkeywords}
%%%%%%%%%%%%%%%%%%%%%%%%%%%%%%%%%%%%%%%%%%%%%%%%%%%%%%%%%%%%%%%%%%%

\section{Introduction}
\label{sec:introduction}

{\subsection{Background and Related Work}}

Nowadays, a growing number of electronic devices and various emerging applications have entered our daily routines, which bring about significant growth in the wireless data traffic of wireless networks and is likely to leap
10000 fold in the next 20 years \cite{GhoshA2014JSAC}-\cite{WangX2018Surveys}.
To tackle this incredible growth, millimeter-wave (mmWave) has become one of the most efficient resolutions due to the plentiful underutilized spectrum resources \cite{HeS2021Proc}. 
Recently, mmWave communication has received substantial attention, such as channel modelling and estimation, beamforming strategy design, and performance analysis.
The authors in \cite{AndrewsJG2017TCOM}  provided a comprehensive overview of mathematical models and analytical techniques of mmWave cellular systems. A baseline mathematical method and analysis in blocking and substantial directionality aspects was proposed and the result indicated that the ultra-dense deployments were more available in mmWave systems. 
In \cite{AlkhateebA2014JSTSP}, an adaptive algorithm was developed to estimate the mmWave channel parameters exploiting the sparse scattering of the channel and was extended to the case of the multi-path channel. 
The result shows that the proposed low-complexity channel estimation algorithm achieves more precoding gains and spectral efficiency than exhaustive algorithms. 
In \cite{LuoZ2021TVT}, a hybrid precoding/combining designs of full-duplex amplify-and-forward mmWave relay systems was proposed. Their simulation indicated that the proposed design is approaching an all-digital scheme. 
A downlink non-orthogonal multiple access (NOMA) mmWave system was investigated in \cite{ZhouY2022TVT}. 
An opportunistic beam-splitting NOMA scheme was proposed to inquire into the antenna gain and the closed-form expressions of the coverage probability and the ASR were derived. %-性能分析

In next-generation communication networks and beyond, interference management becomes a fundamental problem for multi-user communications and multiple access design \cite{ClerckxB2023JSAC}. Rate-splitting multiple access (RSMA) as a candidate multiple access scheme has been proposed in \cite{MaoY2018EURASIP}  and has been recognized as a powerful multiple access, interference management, and multi-user strategy \cite{MaoY2022Surveys}.
Based on the rate-splitting (RS) principle, RSMA not only displays more flexibility in managing interference, i.e., partially decodes interference of inter-user and partially treats interference as noise, but also unifies and generalizes orthogonal multiple access (OMA), space-division multiple access, NOMA, physical-layer multicasting \cite{ClerckxB2020WCL}, \cite{ClerckxB2016CMag}, \cite{ClerckxB2021OJCS}.
A more flexible and powerful cooperative scheme for a multiple-input-single-output (MISO) broadcast channel scenario was proposed based on the three-node relay channel in \cite{ZhangJ2019SPL}. Their simulation results show that the proposed cooperative RS strategy can obtain an explicit rate improvement than NOMA.
In \cite{MaoY2019TCOM}, linearly-precoded 1-layer and multi-layer transmission strategies based on the RS were investigated in a Non-Orthogonal unicast and multicast transmission system, which the weighted sum rate and energy efficiency problems were solved by weighted minimum mean square error algorithm and successive convex approximation algorithm. The Numerical results indicated that the RS-assisted transmission strategies are more efficient in spectrum and energy compared with multi-user linear- precoding, OMA and NOMA in extensive user deployments and network loads.
Compared to rate region, sum rate, spectral and energy efficiency improvement in terms of optimization, the benefit of RSMA for performance analysis, such as throughput, ergodic sum rate (ESR), and outage probability (OP), also needs to be explored.
In \cite{SinghSK2021WCL}, the downlink unmanned aerial vehicle systems based on the RSMA scheme were investigated and the closed-form expressions of the OP and throughput at each user were derived.
The performance of a multi-cell RSMA network was investigated in \cite{ZhuQ2022ArXiv} wherein the analytical expressions for ESR and spectral efficiency based on stochastic geometry theory were derived.
Their results showed that the power splitting ratio between common and private streams significantly impacts performance.
Bansal A. \textit{et al.} in \cite{BansalA2021TVT} studied a novel Intelligent reflecting surface (IRS) RS framework of multi-user communication system, the closed-form expressions of OP for the far and near users were derived. The simulation results showed that the proposed framework was superior to the decode-and-forward-RS, RS without IRS, and IRS-NOMA scenarios.

Physical layer security (PLS), which utilizes the inherent characteristics of wireless channels, is an exciting complement to sophisticated cryptographic techniques \cite{WangN2019IOT}.
Since the mmWave's propagation characteristics,  which support large antenna arrays and highly directional transmission, can reduce the leakage of confidential information. Thus, physical layer security in mmWave communications has attracted considerable recent attention. Specifically, by using a sectored model to analyze the beam pattern,
the secrecy transmission of a mmWave cellular network was investigated in \cite{WangC2016TWC}, the secure connectivity probability was studied and their results indicated the array pattern and intensity of eavesdroppers are both important network parameters for improving the secrecy performance.
The author studied the physical layer security of mmWave relaying networks In \cite{RaghebSM2021TIFS} wherein developed a new joint relay selection and power allocation method in multiple eavesdroppers and relays, the closed-form expressions of average secrecy rate (ASR) and secrecy outage probability (SOP) were derived respectively and demonstrated the superiority of proposed method. 
Ju \textit{et al.} in \cite{JuY2017TCOM} utilized the discrete angular domain channel {{(DADC)}} model to analyze the secrecy performance of the mmWave MISO systems. 
Three transmission schemes were proposed to improve the secure mmWave systems.
Further, they investigated the security mmWave MISO systems in the presence of multiple randomly located eavesdroppers in \cite{JuY2018TWC}.
In \cite{JuY2019TWC},  the multiple input multiple output DADC were developed in mmWave decode-and-forward relay systems.
The secrecy performance with three eavesdropping scenarios was investigated by designing the corresponding beamforming strategies.
Huang \textit{et al.} investigated the secure mmWave NOMA systems in which all the authorized and unauthorized receivers were randomly located in \cite{HuangS2020TVT}.
Then, they proposed two schemes to improve the secrecy transmission and the closed-form SOPs for two beamforming schemes with varying eavesdropper detection abilities were derived.

As a novel, general, and robust framework for the sixth generation mobile communications, RSMA has a high potential to be used for security applications \cite{TedeschiP2020Surveys}.
In \cite{AbolpourM2022OJCS}, two decoding strategies were adopted by a near user and a far user to explore outage and secrecy outage performance. The closed-form expressions of OP and tight approximations of SOP were derived, considering four decoding combinations.
The author in \cite{TongY2022ArXiv} proposed two RS schemes with one-layer-successive interference cancellation (SIC) and two-layer-SIC wherein there is an untrusted near user. The closed-form expressions of OP and SOP are obtained to analyze outage performance.
In \cite{SalemA2022ArXiv}, the secrecy performance of RSMA in multi-user MISO systems was investigated and analytical expressions of the ESR and ASR were obtained with zero-forcing precoding and minimum mean square error approach. The result demonstrated that proposed power allocation methods could offer inherent tradeoffs over the ESR and ASR.

In the RSMA scheme, the messages are split into common and private parts, encoded into a single common stream and private streams, respectively. 
All the common and private streams are superimposed, linearly precoded, and transmitted simultaneously by the transmitter. 
By utilizing SIC technology, the common stream is decoded firstly by treating all private streams as interference, removing from the received signals, and the private stream is decoded by treating the private streams for other users as interference. 
In the mmWave systems with the DADC model, depending on the spatial correlation of the users, the spatially resolvable paths are split into overlapped and non-overlapped paths. 
As thus, the following questions naturally arise: 
\textit{Should the common streams be transmitted on overlapped paths or all paths? What is the performance of the mmWave systems with the RSMA scheme? }
To answer these questions, the performance of the mmWave RSMA systems was investigated, and two beamforming transmission schemes were proposed in \cite{LeiH2023RSMA}. 
The closed-form expressions for the exact and asymptotic OP with proposed schemes were derived using stochastic geometry theory. 
The results demonstrated that RSMA helps improve transmission reliability of the mmWave systems. 
{ TABLE \ref{table11} outlines the typical works discussed.}

\begin{table*}[t]
	\centering
	\caption{Related Literature on RSMA or mmWave systems.}
	\label{table11}
		{
			\begin{tabular}{c|c|c|c|c|c|c}
				\Xhline{1.2pt}{Reference} & {\makecell[c]{RSMA}}  & {mmWave} & {\makecell[c]{Multi-antenna technology}}  & {\makecell[c]{Channel Model}}& PLS &  {\makecell[c]{Performance metric}}\\
				\hline
				\cite{SinghSK2021WCL}   & \checkmark  &   &  & Nakagami-$m$ &    & OP \\
				\hline
				\cite{ZhuQ2022ArXiv}   & \checkmark &  & Transmitter & {\makecell[c]{Rayleigh}} &    & ESR \\
				\hline
				\cite{BansalA2021TVT}   & \checkmark  &  &  & Rayleigh&   &OP \\
				\hline
				\cite{WangC2016TWC}   &  & \checkmark &  &  Nakagami-$m$ & \checkmark  &SCP \\
				\hline
				\cite{RaghebSM2021TIFS}   &  & \checkmark & Transmitter & Rayleigh & \checkmark  &SOP \\
				\hline
				\cite{JuY2017TCOM}   &   & \checkmark & Transmitter &{\makecell[c]{DADC}} & \checkmark  &SOP \\
				\hline
				\cite{JuY2018TWC}   &  & \checkmark  & Transmitter &{\makecell[c]{DADC}} & \checkmark  &SOP \\			
				\hline
				\cite{JuY2019TWC}   &  & \checkmark & {\makecell[c]{Transmitter \& destination}}
				&{\makecell[c]{DADC}} & \checkmark   &SOP \\
				\hline
				\cite{HuangS2020TVT}   &  &  \checkmark & Transmitter &{\makecell[c]{DADC}} & \checkmark   &SOP \\
				\hline
				\cite{AbolpourM2022OJCS}   & \checkmark &  & &Rayleigh & \checkmark   &OP and SOP \\
				\hline
				\cite{TongY2022ArXiv}   & \checkmark &   & Transmitter &Rayleigh & \checkmark   &OP and SOP \\
				\hline
				\cite{SalemA2022ArXiv}   & \checkmark &  & Transmitter &Rayleigh & \checkmark   &{\makecell[c]{ESR \& ASR}} \\
				\hline
				\cite{LeiH2023RSMA}   & \checkmark &  \checkmark & Transmitter & {\makecell[c]{DADC}} &    &OP \\
				\hline
				Our Work   & \checkmark &  \checkmark & Transmitter & {\makecell[c]{DADC}} & \checkmark   &SOP \\
				\Xhline{1.2pt}
		\end{tabular}}
\end{table*}

%%%%%%%%%%%%%%%%%%%%%%%%%%%%%%%%%%%%%%%%%%%%%%%%%%%%%%%%%%%%%%%%%%%
{\subsection{Motivation and Contributions}}

{{To the best of the authors’ knowledge, based on the open literature, there is a lack of research focused on the following issue:
\textit{Is it beneficial to leverage the RSMA scheme in mmWave systems for security enhancement?}
Hence, this work answers this question by analyzing the secrecy performance of the mmWave RSMA systems. Different eavesdropping scenarios are considered corresponding to the overlapped resolvable paths between the main and the wiretap channels under the considered transmission schemes. 
Technically speaking, it is much more challenging to obtain the analytical expression of the SOP than that of the OP for the mmWave RSMA system since there are multiple parameters (random variables) that must be considered.}}
We summarize the contributions of this work as follows.

\begin{enumerate}
	
	\item The secrecy performance of the MISO mmWave RSMA system is investigated wherein the common and private streams are transmitted on the overlapping and non-overlapping paths, respectively. 
	Different scenarios are considered based on the spatial correlation of the legitimate and illegitimate user and the principle of the RSMA scheme. 
	Subsequently, the analytical expressions for the SOP are derived respectively through the Gaussian–Chebyshev quadrature method. 
	
	\item We present Monte Carlo simulation results to validate the correctness of the derived analytical expressions and demonstrate the effects of system parameters on the SOP of the considered mmWave RSMA systems.
	The result shows that the power allocation between the common and private streams and the number of overlapped paths significantly affect the secrecy performance. 
	
	\item Relative to \cite{HuangS2020TVT} wherein the security mmWave NOMA systems were enhanced by the proposed beamforming schemes, the security mmWave RSMA system is improved by the proposed beamforming scheme in this work wherein the scenarios considered in this work are more challenging.
	
	\item {Relative to \cite{AbolpourM2022OJCS}-\cite{LeiH2024WCNC} wherein the secrecy performance of the RSMA systems with {\textit{internal}} untrusted users was investigated. In these scenarios, it was assumed that the common message could always be decoded and only the private message was wiretapped. 
		Technically speaking, it is much more challenging to investigate the secrecy performance of the RSMA systems with an \textit{external} eavesdropper than with an internal eavesdropper.	}
	
\end{enumerate}

\subsection{Organization}
The rest of this paper is organized as follows. Section \ref{sec:SystemModel} describes the system model and beamforming scheme. 
The analytical expressions for the exact SOP of mmWave RSMA systems is derived under different scenarios in Sections \ref{sec:SOPAnalysis}.
The simulation results are presents to valid the analysis in Section \ref{sec:NumericalResults}. 
Section \ref{sec:Conclusion} concludes this work.
{Table \ref{table02} lists the notations and symbols utilized throughout this work.}

\begin{table}[t]
	\centering
	{
		\caption{Notation and Symbols}
		\label{table02}
		\begin{tabular}{l|l}
			\Xhline{1.2pt}
			{Notation} & {Description} \\
			\hline
			${\Omega _l}$          &  The index set of resolvable
			paths at $l$\\
			\hline
			${{\mathbf{g}}_l}$ & The complex gain vector between $S$ and $l$\\
			\hline
			${\mathbf{U}}$ & The spatially orthogonal basis\\
			\hline
			$L_l$               & The number of resolvable paths of $l$\\
			\hline
			${\Delta _{l,o}}$         & The angular range between the $o$th and $(o + 1)$th paths at $l$  \\
			\hline
			${\omega _{l,o}}$    & The width between the $o$th and $(o + 1)$th paths at $l$ \\
			\hline
		    ${\Omega _{ c}}$         & The index set of overlapped paths between $U_1$ and $U_2$ \\
			\hline
			 ${\Omega _{i,p}}$    & The index set of non-overlapping paths at $U_i$ \\
			 \hline
			 ${\Omega _{ec}}$    & The index set of overlapped paths among $E$ and ${\Omega _{ c}}$\\
			 \hline
			 ${\Omega _{e,pi}}$    & The index set of overlapped paths between $E$ and ${\Omega _{i,p}}$\\
			 \hline
			 ${\tau _c}$    &  The power allocation coefficient for $s_c$ \\
			 \hline
			 ${\tau _i}$   &  The power allocation coefficient for $s_i$\\
			 \hline
			 $R_{1,c}^{\mathrm{th}}$    & The secrecy rate threshold for the common messages\\
			 \hline
			 $R_{1,p}^{\mathrm{th}}$     & The secrecy rate threshold for the private messages\\
			 \hline
			$G_{c,d}^{a,b}\left[  \cdot  \right]$  	& Meijer's $G$-function\\
			\hline
			$H_{,:,:,}^{,:,:,}\left[  \cdot  \right]$  & Extended generalized bivariate Fox's $H$-function\\
			\hline
			$\Gamma \left( x \right)$     & Gamma function\\
			\Xhline{1.2pt}
	\end{tabular}}
\end{table}

\section{System Model}
\label{sec:SystemModel}

\subsection{System Model}

\begin{figure}[!t]
	\centering
	\includegraphics[width = 2.5in]{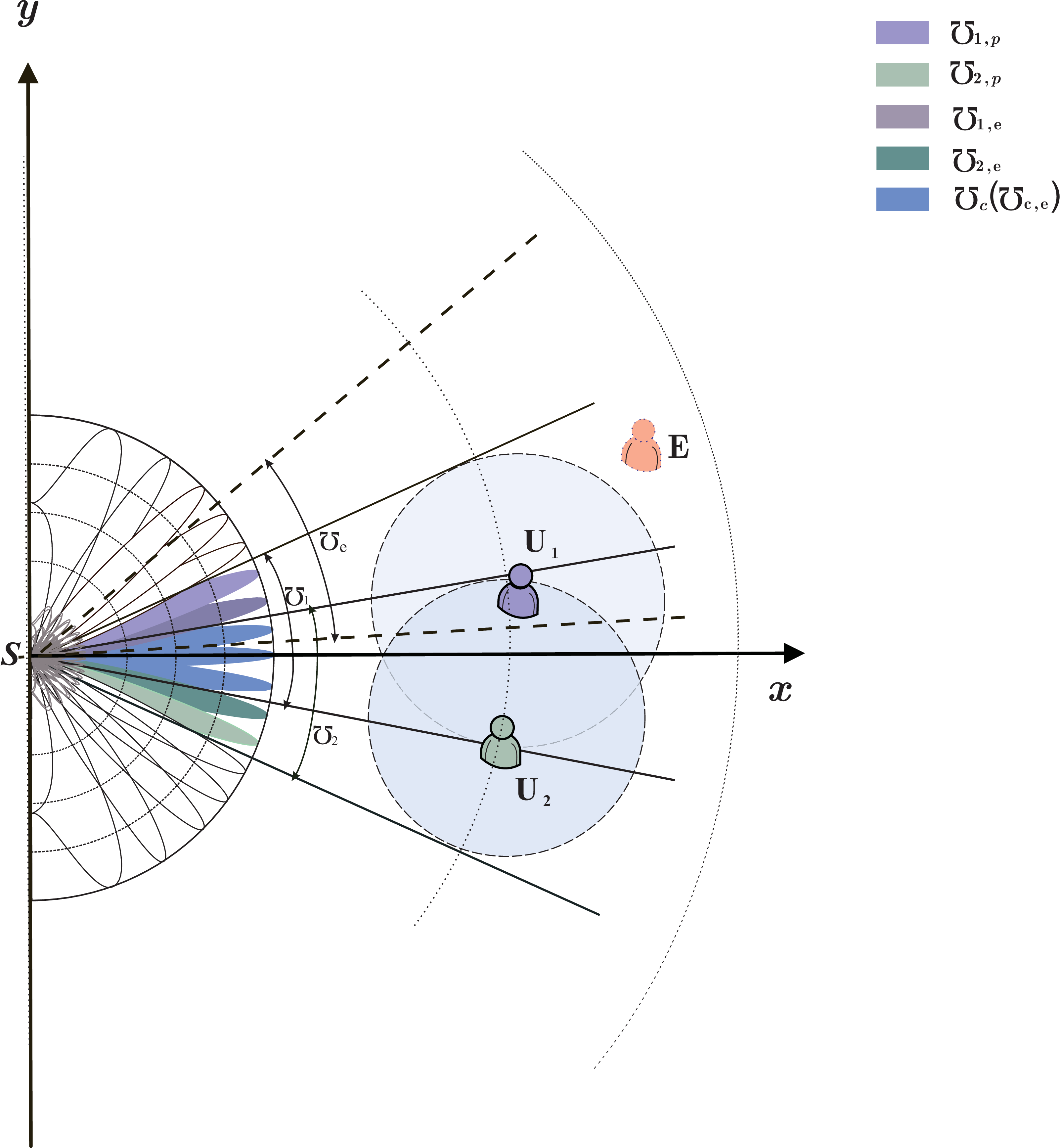}
	\caption{A downlink mmWave RSMA system consists a base station ($S$), two user ($U_1$ and $U_2$), and a external passive eavesdropper ($E$).}
	\label{systemmodel}
\end{figure}

{As shown in Fig. \ref{systemmodel}, a downlink mmWave RSMA system is considered where the base station $\left( S \right)$ communicates with two users denoted by ${U_1}$ and ${U_2}$} \footnote{
	Similar to \cite{MaoY2018EURASIP}, \cite{ClerckxB2020WCL}, \cite{ClerckxB2021OJCS}, the RSMA system with two users is considered in this work. As such, the results in this paper can serve as a benchmark for the performance of such systems. The performance of RSMA systems with multiple users will be part of our future work.	} 
while an external passive eavesdropper denoted by $E$ attempt to intercept the information. 
$S$ is equipped with ${N_s}$ antennas and both legitimate and illegitimate receivers are equipped with a single antenna. 

Similar to \cite{LeiH2023RSMA}, the message, ${W_i}$ $\left( {i = 1,2} \right)$, is split into ${W_{i,c}}$ and ${W_{i,p}}$. 
${W_{1,c}}$ and ${W_{c,2}}$ are encoded together into ${s_{c}}$, which is a common stream decoded by both users.
At the same time, ${W_{1,p}}$ and ${W_{p,2}}$ are encoded into ${s_1}$ and ${s_2}$, respectively.
The transmitted signal from $S$ is given by
\begin{equation}
	{\mathbf{x}} = {{\mathbf{w}}_c}{\sqrt {P{\tau _c}}}{s_c} + {{\mathbf{w}}_1}\sqrt {{{P{\tau _1}}}} {s_1} + {{\mathbf{w}}_2}\sqrt {{{P{\tau _2}}}} {s_2},
	\label{xinhao}
\end{equation}
where
${{\mathbf{w}}_c}$, ${{\mathbf{w}}_1}$, and ${{\mathbf{w}}_2}$ are unit vectors that denote the beamforming direction,
$P$ signifies the transmit power,
and
${\tau _c}$ and ${\tau _i}$ denote the power allocation coefficient for $s_c$ and $s_i$, respectively.

The polar coordinate system is established with the position of $S$ as the origin.
The position of ${U_1}$ and ${U_2}$ are expressed as $\left( {{r_1},{\theta _1}} \right)$ 
and $\left( {{r_2},{\theta _2}} \right)$ respectively, where ${{\theta _i}}$ is the center angle of angles of departure (AODs) of $U_i$'s paths, 
i.e., ${\theta _i} = \frac{{{\theta _{i,\min }} + {\theta _{i,\max }}}}{2}$. 
The channels of between $S$ and the receiver $l$ is expressed as \cite{JuY2017TCOM}-\cite{JuY2019TWC}
\begin{equation}
	{{\mathbf{h}}_{l}} = \sqrt {\frac{{{N_s}P\left( {{r_{l}}} \right)}}{{{L_{l}}}}} {{\mathbf{g}}_{l}}{{\mathbf{U}}^H}, 
	\label{H232}
\end{equation}
where 
$l \in \left\{ {{1},{2}, e} \right\}$, 
$P\left( {{r_l}} \right) = r _0 {r_l}^{ - \alpha }$,
${r_0} = {10^{ - \frac{{{\beta _L}}}{{10}}}}$, %denotes a a frequency-dependent constant,
${\beta _L} = 3.66 + 24.3{\log _{10}}\left( {{f_c}} \right)$,  ${f_c} = 28$ GHz \cite{RappaportST2017TCAP},
$r_l$ denotes the distance between the transmitter and the receiver $l$,
$\alpha $ signifies the path loss exponent,
$L_l$ denotes the number of resolvable paths of the receiver $l$,
${\mathbf{U}}$ is the spatially orthogonal basis,
${{\mathbf{g}}_l} = \left[ {{g_{l,1}},{g_{l,2}}, \ldots ,{g_{l,{N_s}}}} \right]$.
It is assumed that the AODs of the receiver $l$' paths is distributed within the angular range $\left[ {{\theta _{l,\min }},{\theta _{l,\max }}} \right]$, 
where 
${g_{l,n}} \sim CN\left( {0,1} \right)$ if ${\theta _{l,n}} \in \left[ {{\theta _{l,\min }}, {\theta _{l,\max }}} \right]$ otherwise ${g_{l,n}} = 0$ \cite{JuY2017TCOM}. 
To facilitate analysis, we assumed that ${\theta _1} = - {\theta _2} $ and 
${L_1} = {L_{2}} = {L_{e}} = L$. 
We define sets ${\Omega _l} = \left\{ {{I_{l,o}}\left| {{I_{l,o}} \in {Z^ + }} \right.,{\Psi _{{I_{l,o}}}} \in \left[ {\sin \left( {{\theta _{l,\min }}} \right),\sin \left( {{\theta _{l,\max }}} \right)} \right],} \right.{I_{l,1}} < {I_{l,2}}$$\left. { <  \cdots  < {I_{l,L}}} \right\}$. 
So we have 
${{{\mathrm{th}}{\Omega }}_i} = \left\{ {\frac{{{N_{{\mathrm{th}}{s}}} \mp {L_{{\mathrm{th}}{c}}}}}{2} \pm 1,\frac{{{N_{{\mathrm{th}}{s}}} \mp {L_{{\mathrm{th}}{c}}}}}{2} \pm 2, \cdots ,\frac{{{N_{{\mathrm{th}}{s}}} \mp {L_{{\mathrm{th}}{c}}}}}{2} \pm L} \right\}$,
where ${\Omega _{l,o}} = \frac{{{N_s} \mp {L_{c}}}}{2} \pm o$. 
Define the angular range ${\Delta _{l,o}} = \left[ {\arcsin \left( {{\Psi _{\frac{{{N_s} \mp {L_{c}}}}{2} \pm o}}} \right),\arcsin \left( {{\Psi _{\frac{{{N_s} \mp {L_{c}}}}{2} \pm o + 1}}} \right)} \right]$ and the width ${\omega _{l,o}} = \arcsin \left( {{\Psi _{\frac{{{N_s} \mp {L_{c}}}}{2} \pm o + 1}}} \right) - \arcsin \left( {{\Psi _{\frac{{{N_s} \mp {L_{c}}}}{2} \pm o}}} \right)$, where ${\Delta _{l,o}}$ (${1 \le o < L}$) describes the angular range  of user $l$ between the $o${th} and $\left( {o + 1} \right)$th spatially resolvable paths, ${\omega _{l,o}}$ denote the width between the $o${th} and $\left( {o + 1} \right)$th spatially resolvable paths..
Then we have 
$\left| {{\Omega _1}} \right| = \left| {{\Omega _2}} \right| = \left| {{\Omega _e}} \right|  = L$.

Based on the spatial correlation of all receivers, the spatially resolvable paths are divided into overlapped and non-overlapping paths \cite{JuY2017TCOM}.
${L_c}$ and $L_p$ are defined as the number of the overlapped and non-overlapping paths between $U_1$ and $U_2$, respectively.  
Thus, we have ${\Omega _{ c}} = {\Omega _1} \cap {\Omega _2}$,
${\Omega _{1,p}} = {\Omega _1} - {\Omega _{c}}$, 
${\Omega _{2,p}} = {\Omega _2} - {\Omega _{c}}$.
$\left| {{\Omega _c}} \right| = {L_c}$, 
and 
$\left| {{\Omega _{1,p}}} \right| = \left| {{\Omega _{2,p}}} \right| = {L_p}$.
Similarly, ${L_{ec}}$, ${L_{e1}}$, ${L_{e2}}$ denote the number of the overlapped paths between ${\Omega _{e}}$, 
and 
${\Omega _{c}}$, ${\Omega _{1,p}}$,  ${\Omega _{2,p}}$, respectively.
%${L_c}$ and $E$ and ${L_{e1}}$ denote the overlapped paths between the non-overlapping paths of $U_i$ and $E$.
Thus, we have 
%${\Omega _{1, c}} = {\Omega _1} \cap {\Omega _2}$, 
${\Omega _{ec}} = {\Omega _{c}} \cap {\Omega _e}$ 
and ${\Omega _{e, p1}} = {\Omega _{1, p}} \cap {\Omega _e}$, ${\Omega _{e, p2}} = {\Omega _{2, p}} \cap {\Omega _e}$,
$\left| {{\Omega _{e,c}}} \right| = {L_{ec}}$, 
$\left| {{\Omega _{e,p1}}} \right| = {L_{e1}}$, 
$\left| {{\Omega _{e,p2}}} \right| = {L_{e2}}$.

In this work, $s_c$ is transmitted on the overlapped paths and $s_i$ is transmitted on their non-overlapping paths to eliminate the inter-user interference of private signals by utilizing the spatial correlation of two users' channels \cite{LeiH2023RSMA}. 
Similar to \cite{JuY2017TCOM} and \cite{LeiH2023RSMA}, 
{{it is assumed that the perfect channel state information (CSI) of the legitimate receivers is available at $S$ }},
the beamforming for $s_c$ is expressed as ${{\mathbf{w}}_{c}} = {\cal S}\left( {{\mathbf{U}},{\Omega _{ c}}} \right){ \in ^{{N_s} \times {L_{c}}}}$, 
where 
${\cal S}\left( {{\mathbf{B}},{D}} \right)$ is utilized to generate a new matrix with columns selected from ${\mathbf{B}}$ based on the selected column index set {${D}$},
and ${{\Omega _{c}}}$ denotes the index sets of common resolvable paths between the receiver ${U_1}$ and ${U_2}$.
Similarly, the beamforming for ${s_i}$ is expressed as $	{{\mathbf{w}}_i} = S\left( {{\mathbf{U}},{\Omega _{i, p}}} \right){ \in ^{{N_s} \times {L_p} }}$, 
where ${{\Omega _{i, p}}}$ denotes the index set of the non-overlapping paths of receiver ${U_i}$. Thus, the transmitted signal from $S$ is rewritten as
\begin{equation}
	{\mathbf{x}} = {{\mathbf{w}}_c}\sqrt {P{\tau _c}} {{\mathbf{s}}_c} + {{\mathbf{w}}_1}\sqrt {P{\tau _1}} {{\mathbf{s}}_1} + {{\mathbf{w}}_2}\sqrt {P{\tau _2}} {{\mathbf{s}}_2},
	\label{xinhao}
\end{equation}
where ${{\mathbf{s}}_c} \in {{\mathbb{C}}^{{L_c} \times 1}}$ and ${{\mathbf{s}}_i} \in {{\mathbb{C}}^{{L_p} \times 1}}$.

\subsection{Signal to Interference Plus Noise Ratio}

According to the RSMA principle, ${U_i}$ decodes ${s_c}$ firstly by treating all the other signals as noise. 
Then the instantaneous signal-to-interference and noise ratio (SINR) of decoding ${s_c}$ at ${U_i}$ is 
${\gamma _{i, c}} = \frac{{{\delta _{i,c}}{{\left\| {{{\mathbf{g}}_{i,c}}} \right\|}^2}}}{{{\delta _i}{{\left\| {{{\mathbf{g}}_{{i, p}}}} \right\|}^2} + 1}}$, 
where
${\delta _{i, c}} = \frac{{\delta {\tau _{c}}}}{{r_i^\alpha }}$,
${\delta _i} = \frac{{\delta {\tau _i}}}{{r_i^\alpha }}$,
$\rho  = \frac{P}{{\sigma ^2}}$, 
and
$\delta  = \frac{{{N_s}\rho {r _0} }}{L}$.
After performing SIC, i.e., $s_c$ is re-encoded, precoded, and removed from the received signal, the SINR of decoding $s_i$ at ${U_i}$ is obtained as 
${\gamma _{i, p}} = {\delta _i}{\left\| {{{\mathbf{g}}_{{i, p}}}} \right\|^2}$.

\begin{figure}[t]
	\centering
	\subfigure[${\theta _{e,\min }} \in {\Delta _{L - {L_{e1}}}}$. ]{
		\label{fig1a}
		\includegraphics[width = 0.231 \textwidth]{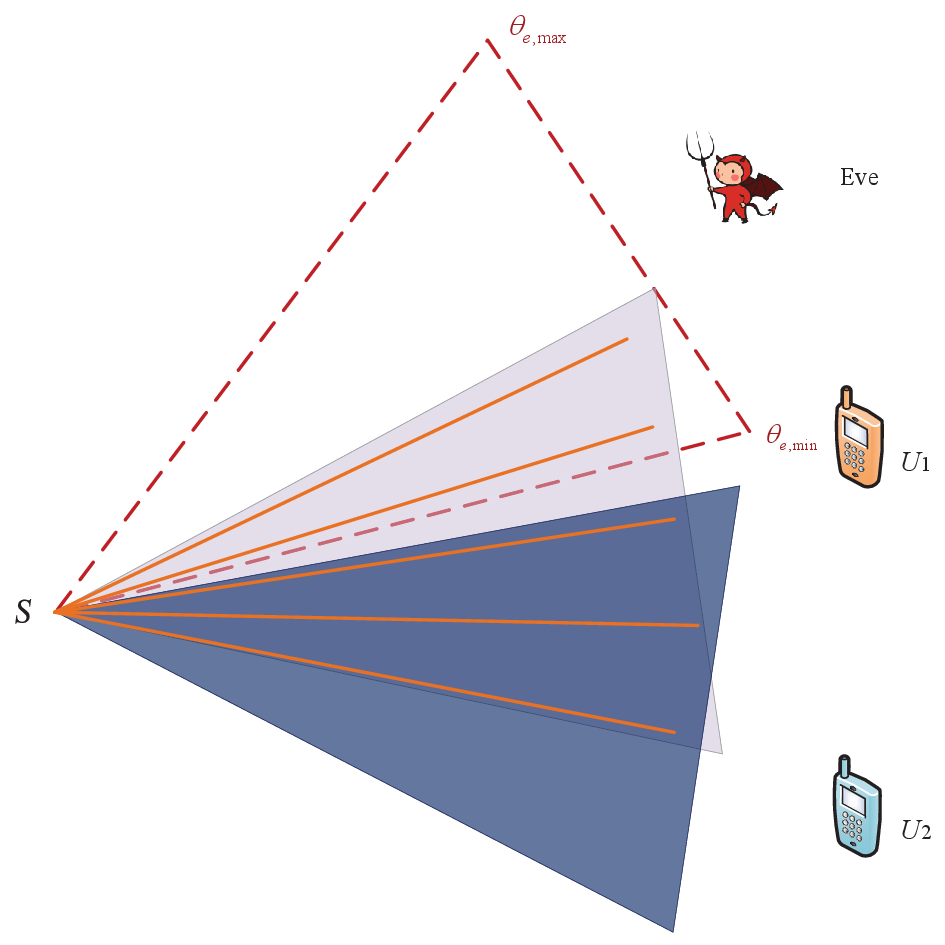}}
	\subfigure[${\theta _{e,\min }} \in {\Delta _{L - {L_{p}} - {L_{ec}}}}$. ]{
		\label{fig1b}
		\includegraphics[width = 0.231 \textwidth]{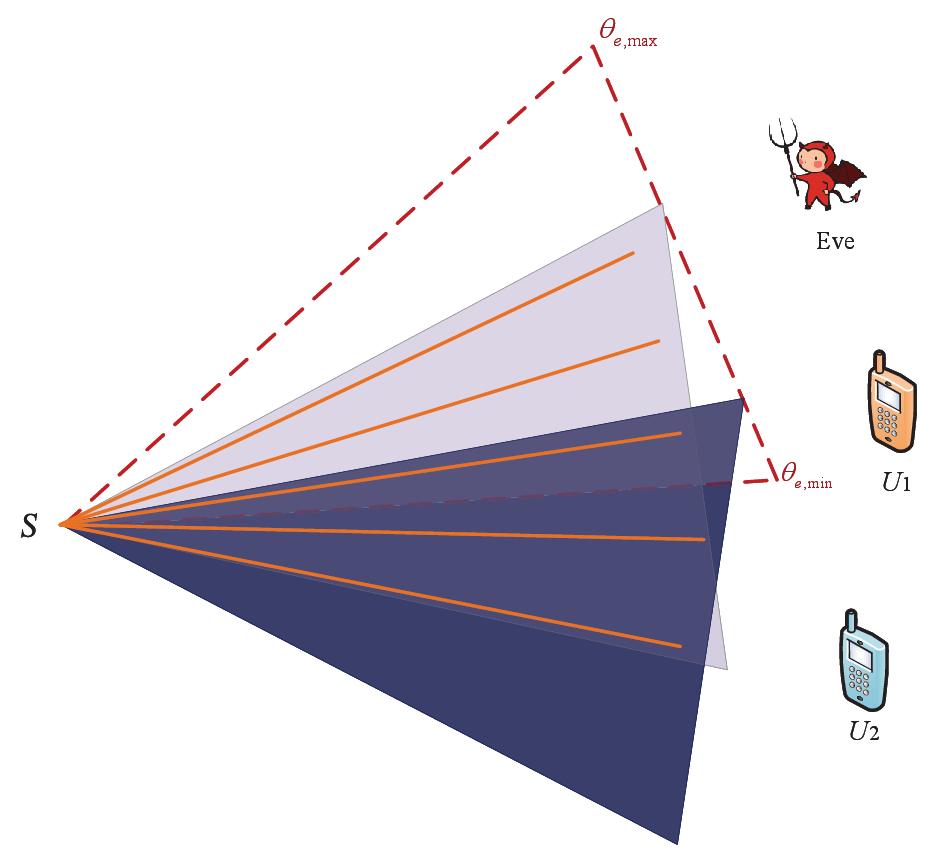}}
	\subfigure[${\theta _{e,\max }} \in {\Delta _{{L_{ec}} + {L_{e1}}}}$. ]{
		\label{fig1c}
		\includegraphics[width = 0.231 \textwidth]{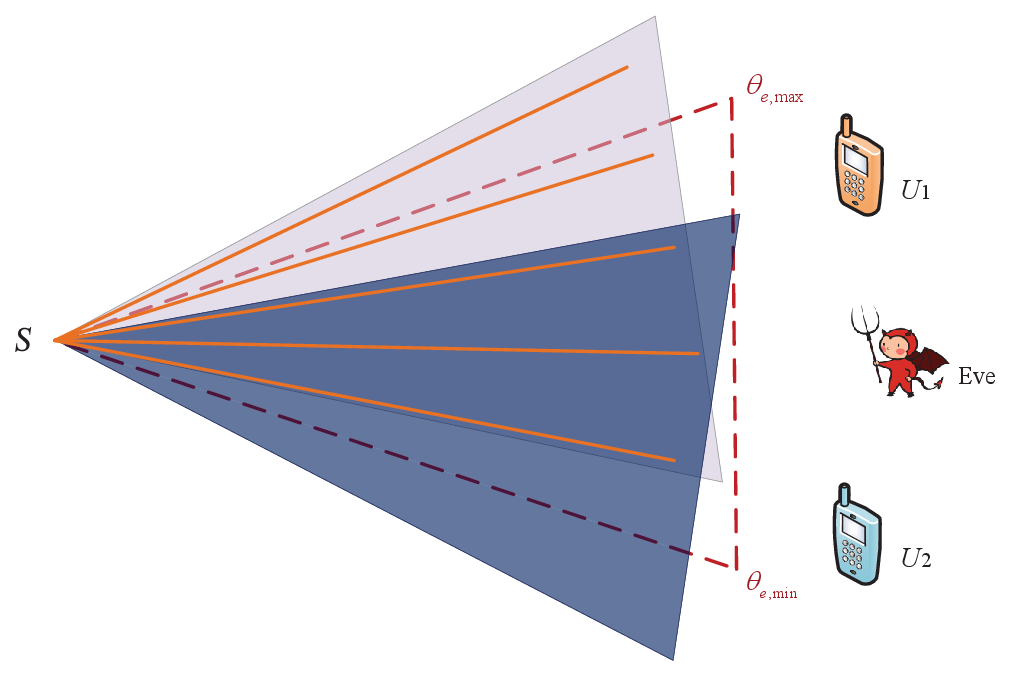}}
	\subfigure[${\theta _{e,\max }} \in {\Delta _{{L_{ec}}}}$. ]{
		\label{fig1d}
		\includegraphics[width = 0.231 \textwidth]{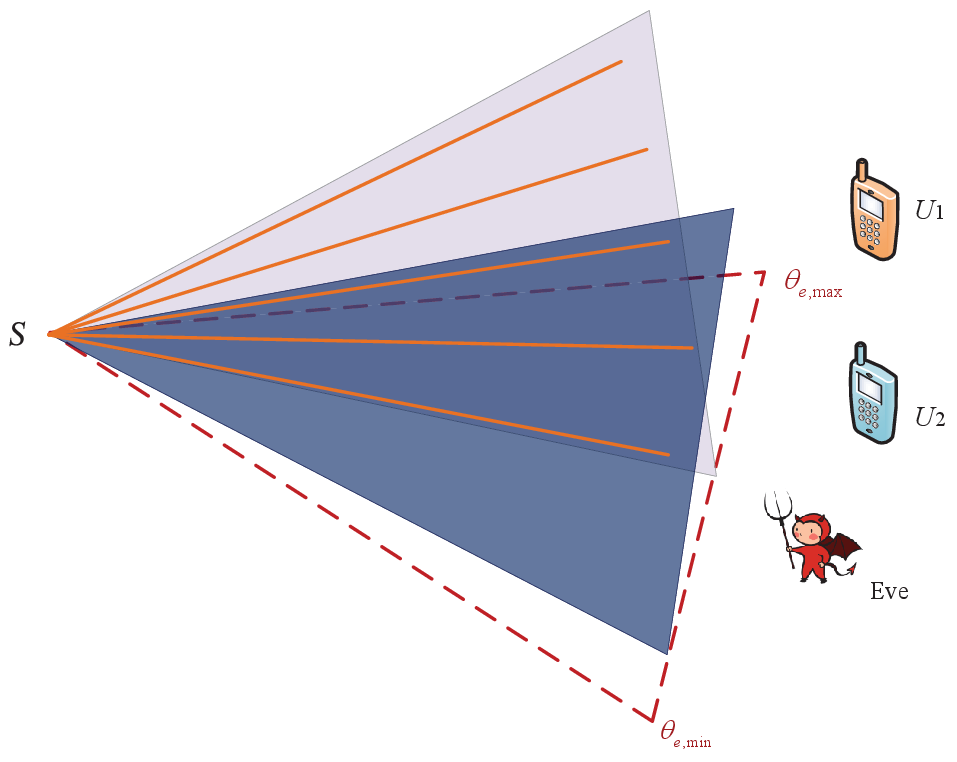}}
	\caption{Scenarios for the different number of the overlapped overlapped paths between $U_i$ and $E$ with $L = 5$ and ${L_{c}} = 3$.}
	\label{fig1}
\end{figure}

In this work, it is assumed that $E$ is interested only in $U_1$'s message 
\footnote{ 
	Since the analysis for both users is interchangeable, the secrecy performance of the scenarios in which $E$ is interested in $U_2$'s message can be obtained based on the results of this manuscript. 
}. 
Moreover, for tractability of analysis, 
${\Delta _{1,o}}$ (${1 \le o < L}$) is represented by ${\Delta _{o}}$,
the users' locations are assumed to be fixed in this work \footnote{
	The secrecy performance of scenarios wherein the users are randomly distributed also can be easily obtained based on \cite[(7)]{LeiH2023RSMA} and the results of this work. 
}. 
Depending on the relative positions of $U_1$ and $E$, as shown in Fig. \ref{fig1}, there are four scenarios for the SINR of $E$ \footnote{
{Except for the four scenarios considered here, there is another scenario wherein there is no overlapped path between $E$ and $U_1$ and both the common and private messages are secure. }
}, 
{ where ${\Delta _{L - {L_{e1}}}}$ signifies the angular range of $U_1$ between the $\left( {L-{L_{e1}}} \right)$th and $\left( {L-{L_{e1}} + 1} \right)$th spatially resolvable paths, similarly, ${\Delta _{L - {L_p} - {L_{ec}}}}$ signifies the angular range of $U_1$ between the $\left( {L - {L_p} - {L_{ec}}} \right)$th and $\left( {L - {L_p} - {L_{ec}}} + 1 \right)$th spatially resolvable paths.}

\textbf{1) Scenario I}: ${\theta _{e,\min }} \in {\Delta _{L - {L_{e1}}}}$

As show in Fig. \ref{fig1a}, only the private stream of $U_1$ is wiretapped by $E$, which denotes ${\gamma _{e,c}^{\mathrm{I}}} = 0$. 
The SINR of $E$ in eavesdropping private stream of $U_1$ is given as ${\gamma _{e,p1}^{\mathrm{I}}} = {\delta _{e1}}{\left\| {{{\mathbf{g}}_{e,p1}}} \right\|^2},$
where 
${\delta _{e1}} = \delta {\tau _1}r_e^{ - \alpha }$.

\textbf{{2) Scenario II}}: ${\theta _{e,\min }} \in {\Delta _{L - {L_{p}} - {L_{ec}}}}$

This scenario is shown in Fig. \ref{fig1b}, where in both common and private streams of $U_1$ are eavesdropped by $E$. 
More Specifically, all non-overlapping  resolvable paths of $U_1$ is wiretapped and overlapped paths of $U_1$ is partially wiretapped. 
{ 
	Like \cite{LeiH2020TCOM}, it is assumed that illegitimate has the same decoding capability as the legitimate users\footnote{
		
%		There are some outstanding works, such as \cite{AbolpourM2022OJCS}-\cite{LeiH2024WCNC}, investigated the secrecy performance of the RSMA systems with {\textit{internal}} untrusted users. In these scenarios, it is assumed that the common message can always be decoded and only the private message is wiretapped. 
%		Technically speaking, it is much more challenging to investigate the secrecy performance of the RSMA systems with an external eavesdropper than with an internal eavesdropper. 
%		Moreover, t
		There is another eavesdropping scenario, called the worst-case security scenario, considered in many works, such as \cite{LeiH2020TCOM, Lei2024TWC}. Specifically, by utilizing specific detection techniques, the data stream received can be distinguished by the eavesdropper by subtracting interference generated by the superposed signals from each other. This assumption has been utilized in a reasonable amount of literature focused on the physical layer security of NOMA systems. In fact, this assumption simplified the analysis but overestimated the eavesdropper's multi-user decodability. 
		}.
	According to the RSMA principle, $E$ decodes common streams of $U_1$ firstly by treating all the other signals as noise, and then decodes private streams of $U_1$ }.
Thus, the SINR of $E$ in eavesdropping both the common stream and private stream of $U_1$ are given as
\begin{equation}
	{\gamma _{e,c}^{\mathrm{II}}} = \frac{{{\delta _{ec}}{{\left\| {{{\mathbf{g}}_{e,c}}} \right\|}^2}}}{{{\delta _{e1}}{{\left\| {{{\mathbf{g}}_{e,p1}}} \right\|}^2} + 1}},
	\label{SNREc_2}
\end{equation}
\begin{equation}
	{\gamma _{e,p1}^{\mathrm{II}}} = {\delta _{e1}}{\left\| {{{\mathbf{g}}_{e,p1}}} \right\|^2},
	\label{SNREp1_2}
\end{equation}
respectively,
where
${\delta _{ec}} = \delta {\tau _{c}}r_e^{ - \alpha }$.

\textbf{{3) Scenario III}}: ${\theta _{e,\max }} \in {\Delta _{{L_{ec}} + {L_{e1}}}}$

As show in Fig. \ref{fig1c}, all the overlapped paths and part of the non-overlapping paths for  $U_1$ is intercepted. 
{ Similarly, the $E$ firstly decodes common streams and then decodes private streams of $U_1$}.
Thus, the SINR of $E$ in eavesdropping the common stream and private stream of $U_1$ are given as
\begin{equation}
	{\gamma _{e,c}^{\mathrm{III}}} = \frac{{{\delta _{ec}}{{\left\| {{{\mathbf{g}}_{e,c}}} \right\|}^2}}}{{{\delta _{e1}}{{\left\| {{{\mathbf{g}}_{e,p1}}} \right\|}^2} + {\delta _{e2}}{{\left\| {{{\mathbf{g}}_{e,p2}}} \right\|}^2} + 1}},
	\label{SNREci_3}
\end{equation}
and 
\begin{equation}
	{\gamma _{e,p1}^{\mathrm{III}}} = \frac{{{\delta _{e1}}{{\left\| {{{\mathbf{g}}_{e,p1}}} \right\|}^2}}}{{{\delta _{e2}}{{\left\| {{{\mathbf{g}}_{e,p2}}} \right\|}^2} + 1}},
	\label{SNREpi_3}
\end{equation}
respectively,
where
${\delta _{e2}} = \delta {\tau _2}r_e^{ - \alpha }$.

\textbf{4) Scenario IV}: ${\theta _{e,\max }} \in {\Delta _{{L_{ec}}}}$

This case is shown in Fig. \ref{fig1d} wherein only the common stream of $U_1$ is eavesdropped by $E$, which denotes ${\gamma _{e,p1}^{\mathrm{IV}}} =0$ and 
\begin{equation}
	{\gamma _{e,c}^{\mathrm{IV}}} = \frac{{{\delta _{ec}}{{\left\| {{{\mathbf{g}}_{e,c}}} \right\|}^2}}}{{{\varpi _0}{\delta _{e2}}{{\left\| {{{\mathbf{g}}_{e,p2}}} \right\|}^2} + 1}}.
	\label{SNRe_4}
\end{equation}

To facilitate analysis, we define
${X _{lq}} = {\left\| {{{\mathbf{g}}_{l,q}}} \right\|^2}$ 
$\left( {\left( {l,q} \right) \in \left\{ {\left( {i,c} \right),\left( {i,p} \right),\left( {e,c1} \right),\left( {e,p1} \right),\left( {e,p2} \right)} \right\}} \right)$.
The PDF and CDF of ${X _{l,q}} $ are expressed as
\begin{equation}
	{f_{{X_{lq}}}}\left( x \right) = \frac{{{e^{ - x}}{x^{{\kappa _{lq}} - 1}}}}{{\Gamma \left( {{\kappa _{lq}}} \right)}},
	\label{xpdf}
\end{equation}
and
\begin{equation}
	{F_{{X_{lq}}}}\left( x \right) = 1 - {e^{ - x}}\sum\limits_{t = 0}^{{\kappa _{lq}} - 1} {\frac{{{x^t}}}{{t!}}},
	\label{xcdf}
\end{equation}
respectively,
where ${\kappa _{1c}} = {\kappa _{2c}} = {L_c}$,
${\kappa _{1p}} = {\kappa _{2p}} = {L_p}$, 
${\kappa _{ec1}} = {L_{ec}}$, 
${\kappa _{ep1}} = {L_{e1}}$, 
and 
${\kappa _{ep2}} = {L_{e2}}$.

\section{Secrecy Outage Probability Analysis}
\label{sec:SOPAnalysis}

In this work, $U_1$ is secure only when both the common and private stream are secrecy. 
Thus, the SOP of $U_1$ in the $j$th $\left( {j \in \left\{ {{\mathrm{I}},{\mathrm{II}},{\mathrm{III}},{\mathrm{IV}}} \right\}} \right)$ scenario is expressed as
\begin{equation}
	{P_{{\mathrm{sop}},1}^j} = 1 - \underbrace {\Pr \left\{ {C_{1,c}^{{\mathrm{s}},j} > R_{1,c}^{{\mathrm{th}}},C_{1,p}^{\mathrm{s},j} > R_{1,p}^{{\mathrm{th}}}} \right\}}_{ \buildrel \Delta \over = {P_{{\mathrm{scp}},1}^j}},
	\label{sopi}
\end{equation}
where 
${C_{1,c}^{\mathrm{s},j}} = {\left[ {{{\log }_2}\left( {1 + {\gamma _{1,c}}} \right) - {{\log }_2}\left( {1 + {\gamma _{e,c}^j}} \right)} \right]^ + }$ denotes instantaneous secrecy capacity of common streams intended to $U_1$,
${C_{1,p}^{\mathrm{s},j}} = {\left[ {{{\log }_2}\left( {1 + {\gamma _{1,p}}} \right) - {{\log }_2}\left( {1 + {\gamma _{e,p1}^j}} \right)} \right]^ + }$ denotes instantaneous secrecy capacity of private streams transmitted to the $U_1$, 
${\left[ x \right]^ + } = \max \left\{ {x,0} \right\}$,
and 
{ $R_{1,c}^{\mathrm{th}}$ and $R_{1,p}^{\mathrm{th}}$ signify the secrecy rate threshold for the common and private messages, respectively,}
and
${P_{{\mathrm{scp}},1}^j}$ denotes the secrecy connection probability (SCP), which is the complementary of SOP.
It should be noted that, when $E$ can only wiretap the common streams or private streams, the SCP of $U_1$ is degenerated to
${P_{{{\mathrm{scp}}},1}^j} = \Pr \left\{ {{C_{1,c}^{{\mathrm{s}},j}} > {R_{1,c}^{\mathrm{th}}}} \right\}$ or ${P_{{{\mathrm{scp}}},1}^j} = \Pr \left\{ {{C_{1,p}^{{\mathrm{s}},j}} > {R_{1,p}^{\mathrm{th}}}} \right\}$ respectively. 

\textbf{1) Scenario I}: ${\theta _{e,\min }} \in {\Delta _{L - {L_{e1}}}}$

In this case, $E$ only eavesdrop on private streams of $U_1$.
Based on (\ref{xpdf}), (\ref{xcdf}) and (\ref{sopi}), utilizing \cite[(3.351.3)] {Gradshteyn2007Book}, ${P_{{\mathrm{scp}},1}^{\mathrm{I}}}$ is obtained as
\begin{equation}
	\begin{aligned}
		{P_{{\mathrm{scp}},1}^{\mathrm{I}}} &= \Pr \left\{ {{{\log }_2}\left( {\frac{{1 + {\gamma _{1,p}}}}{{1 + {\gamma _{e,p1}^{\mathrm{I}}}}}} \right) > {R_{1,p}^{\mathrm{th}}}} \right\}\\
		%&= \Pr \left\{ {\frac{{1 + {\delta _1}{X_{1p}}}}{{1 + {\delta _{e1}}{X_{ep1}}}} > {\Theta _{1,p}}} \right\}\\
		&= \Pr \left\{ {{X_{1p}} > {A_1}{X_{ep1}} + {A_2}} \right\}\\
		&= {{\mathbb{E}}_{{X_{ep1}}}}\left[ {{e^{ - \left( {{A_1}{X_{ep1}} + {A_2}} \right)}}\sum\limits_{t = 0}^{{L_p} - 1} {\frac{{{{\left( {{A_1}{X_{ep1}} + {A_2}} \right)}^t}}}{{t!}}} } \right]\\
		&= \sum\limits_{t = 0}^{{L_p} - 1} {\sum\limits_{n = 0}^t {\frac{{{e^{ - {A_2}}}A_1^nA_2^{t - n}}}{{n!\left( {t - n} \right)!\left( {{L_{e1}} - 1} \right)!}}} }\\
		& \times\int_0^\infty  {{e^{ - \left( {{A_1} + 1} \right)x}}} {x^{{L_{e1}} + n - 1}}dx\\
		&= \sum\limits_{t = 0}^{{L_p} - 1} {\sum\limits_{n = 0}^t {\frac{{{e^{ - {A_2}}}A_1^nA_2^{t - n}\left( {{L_{e1}} + n - 1} \right)!}}{{n!\left( {t - n} \right)!\left( {{L_{e1}} - 1} \right)!{{\left( {{A_1} + 1} \right)}^{{L_{e1}} + n}}}}} },
	\end{aligned}
	\label{SCP_p1_medi1}
\end{equation}
where 
${\mathbb{E}}\left[ . \right]$ denotes the expectation operation, 
${\Theta _{1,p}} = {2^{{R_{1,p}^{\mathrm{th}}}}}$,
${A_1} = \frac{{{\Theta _{1,p}}{\delta _{e1}}}}{{{\delta _1}}} = \frac{{{\Theta _{1,p}}r_1^\alpha }}{{r_e^\alpha }}$, 
and
${A_2} = \frac{{{\Theta _{1,p}} - 1}}{{{\delta _1}}}$.

\textbf{{2) Scenario II}}: ${\theta _{e,\min }} \in {\Delta _{L - {L_{p}} - {L_{ec}}}}$

In this case, the private streams is eavesdropped completely while the common streams is partly eavesdropped.
Thus, the SCP of $U_1$ is expressed as
\begin{equation}
	\begin{aligned}
			{P_{{\mathrm{scp}},1}^{\mathrm{II}}} &=\Pr \left\{ {{{\log }_2}\left( {\frac{{1 + {\gamma _{1,c}}}}{{1 + {\gamma _{e,c}^{\mathrm{II}}}}}} \right) > {R_{1,c}^{\mathrm{th}}},} \right.\\
			&\left. {{{\log }_2}\left( {\frac{{1 + {\gamma _{1,p}}}}{{1 + {\gamma _{e,p1}^{\mathrm{II}}}}}} \right) > {R_{1,p}^{\mathrm{th}}}} \right\}\\
			& = \Pr \left\{ {\frac{{1 + \frac{{{\delta _{1,c}}{X_{1c}}}}{{{X_1}}}}}{{1 + \frac{{{\delta _{ec}}{X_{ec}}}}{{{X_2}}}}} > {\Theta _{1,c}},\frac{{{X_1}}}{{{X_2}}} > {\Theta _{1,p}}} \right\}\\
			&= \int_1^\infty  {\int_{{{X}_2}{\Theta _{1,p}}}^\infty  {{\Delta _1}{f_{{{X}_1}}}\left( {{{X}_1}} \right)d{{X}_1}{f_{{{X}_2}}}\left( {{{X}_2}} \right)d{{X}_2}} },
	\end{aligned}
	\label{P_scp_2_medium}
\end{equation}
where 
${\Delta _1} = \int_{{{X}_1}{\eta _1}}^\infty  {{F_{{X_{ec}}}}\left( {\left( {\frac{y}{{{{X}_1}{\eta _2}}} - \frac{{{\eta _1}}}{{{\eta _2}}}} \right){{X}_2}} \right){f_{{X_{1c}}}}\left( y \right)dy} $, 
${{X}_1} = 1 + {\delta _1}{X_{1p}}$, 
${{X}_2} = 1 + {\delta _{e1}}{X_{ep1}}$, 
${\eta _1} = \frac{{{\Theta _{1,c}} - 1}}{{{\delta _{1, c}}}}$, 
${\eta _2} = \frac{{{\Theta _{1,c}}{\delta _{ec}}}}{{{\delta _{1, c}}}}$.
Based on  (\ref{xpdf}) and (\ref{xcdf}) and 
utilizing \cite[(3.351.2)]{Gradshteyn2007Book}, ${\Delta _1}$ is obtained as 
\begin{equation}
	\begin{aligned}
				{\Delta _1} &= {{\bar F}_{{X_{1c}}}}\left( {{{X}_1}{\eta _1}} \right) \\
				& - \sum\limits_{t = 0}^{{L_{ec}} - 1} {\sum\limits_{m = 0}^t {\sum\limits_{k = 0}^{m + {L_c} - 1} {\frac{{{A_3}{X}_1^{{L_c}}{X}_2^t{e^{ - {{X}_1}{\eta _1}}}}}{{{{\left( {{{X}_2} + {{X}_1}{\eta _2}} \right)}^{m + {L_c} - k}}}}} } },
	\end{aligned}
\label{Delta_1}
\end{equation}
where
${A_3} = \frac{{{{\left( { - 1} \right)}^{t - m}}\left( {m + {L_c} - 1} \right)!{\eta _1}^{t - m + k}\eta _2^{m + {L_c} - k - t}}}{{k!m!\left( {t - m} \right)!\Gamma \left( {{L_c}} \right)}}$ 
and
${{\bar F}_{{X_{1c}}}}\left( x \right) = 1 - {F_{{X_{1c}}}}\left( x \right)$. 
Define ${X_I} = 1 + \varepsilon X$, $\left( {I = 1,2} \right)$, %based on (\ref{xcdf}), 
the PDF of ${X_I}$ is obtained as 
\begin{equation}
	{f_{{X_I}}}\left( x \right) = \frac{{{e^{ - \frac{{x - 1}}{\varepsilon }}}{{\left( {x - 1} \right)}^{{\kappa _X} - 1}}}}{{{\varepsilon ^{{\kappa _X}}}\left( {{\kappa _X} - 1} \right)!}}.
	\label{X_PDF}
\end{equation}
Then, 
${P_{{\mathrm{scp}},1}^{\mathrm{II}}}$ is obtained as
\begin{equation}
	\begin{aligned}
		{P_{{\mathrm{scp}},1}^{\mathrm{II}}} &= \sum\limits_{t = 0}^{{L_c} - 1} {\frac{{\delta _1^{ - {L_p}}\eta _1^t\delta _{e1}^{ - {L_{e1}}}{\Delta _2}}}{{t!\left( {{L_p} - 1} \right)!\left( {{L_{e1}} - 1} \right)!}}} \\
				& - \sum\limits_{t = 0}^{{L_{ec}} - 1} {\sum\limits_{m = 0}^t {\sum\limits_{k = 0}^{m + {L_c} - 1} {\sum\limits_{n = 0}^{{L_c}} {A_3}{A_4}{\Delta _3} } } },
	\end{aligned}
	\label{scp_2_medium2}
\end{equation}
where
${A_4} = \frac{{\left( {{L_c}} \right)!{e^{ - {\eta _1}}}\delta _1^{ - {L_p}}\delta _{e1}^{ - {L_{e1}}}}}{{n!\left( {{L_c} - n} \right)!\left( {{L_p} - 1} \right)!\left( {{L_{e1}} - 1} \right)!}}$,
${\Delta _2} = \int_1^\infty  {\int_{y{\Theta _{1,p}}}^\infty  {{e^{ - x{\eta _1}}}{x^t}{e^{ - \frac{{x - 1}}{{{\delta _1}}}}}{{\left( {x - 1} \right)}^{{L_p} - 1}}dx} } {e^{ - \frac{{y - 1}}{{{\delta _{e1}}}}}}{\left( {y - 1} \right)^{{L_{e1}} - 1}}dy$
and
${\Delta _3} = \int_1^\infty  {\int_{y{\Theta _{1,p}} - 1}^\infty  {\frac{{{e^{ - \left( {{\eta _1} + \frac{1}{{{\delta _1}}}} \right)z}}{z^{n + {L_p} - 1}}dz}}{{{{\left( {z{\eta _2} + y + {\eta _2}} \right)}^{m + {L_c} - k}}}}} \frac{{{y^t}{{\left( {y - 1} \right)}^{{L_{e1}} - 1}}dy}}{{{e^{\frac{{y - 1}}{{{\delta _{e1}}}}}}}}} $.

Utilizing \cite[(3.351.2), (3.351.3)]{Gradshteyn2007Book}, ${\Delta _2}$ is obtained as (\ref{delta_2}), shown at the top of this page, where
${A_5} = \frac{{\left( {m + {L_p} - 1} \right)!t!{e^{\frac{1}{{{\delta _1}}} - {\Theta _{1,p}}\left( {{\eta _1} + \frac{1}{{{\delta _1}}}} \right)}}{{\left( { - 1} \right)}^{k - i}}\Theta _{1,p}^i}}{{m!\left( {t - m} \right)!{{\left( {{\eta _1} + \frac{1}{{{\delta _1}}}} \right)}^{m + {L_p} - k}}\left( {k - i} \right)!j!\left( {i - j} \right)!}}$.
Similarly, based on (\ref{scp_2_medium2}) and utilizing \cite[(7.811.5)]{Gradshteyn2007Book} and (25.4.39), we obtain ${\Delta _3}$ as (\ref{delta_3}), shown at the top of next page,
where
${A_6} = \frac{{\pi \sqrt {{b_1} - b_1^2} t!\left( {n + {L_p}} \right)!\Theta _{1,p}^j{{\left( {{\Theta _{1,p}} - 1} \right)}^{n + {L_p} - j}}}}{{Ib_1^{{a_3}}q!\left( {t - q} \right)!j!\left( {n + {L_p} - j!} \right){e^{\left( {\frac{{{\eta _1}}}{{{b_1}}} + \frac{1}{{{b_1}{\delta _1}}}} \right)\left( {{\Theta _{1,p}} - 1} \right)}}}}$,
${b_1} = \frac{1}{2}\left( {1 + \cos \frac{{\left( {2i - 1} \right)\pi }}{{2I}}} \right)$, 
$G_{p,q}^{m,n}\left[ \cdot \right]$ is the Meijer's $G$-function as defined by \cite[(9.301)]{Gradshteyn2007Book},
$I$ is the summation items that reflect accuracy versus complexity,
${a_1} = n + {L_p} + 1 - m - {L_c} + k$,
${a_2} = j + q + {L_{e1}} - m - {L_c} + k$,
${b_2} = {\eta _2}\left( {{\Theta _{1,p}} - 1} \right) + {b_1} + {b_1}{\eta _2}$
and
${b_3} = \frac{{{b_2}}}{{{\eta _2}{\Theta _{1,p}} + {b_1}}}$.
\newcounter{TempEqCnt2}
\setcounter{TempEqCnt2}{\value{equation}}
\setcounter{equation}{16} %指定跨栏公式的序号
\begin{figure*}[ht]
	\begin{equation}
	\begin{aligned}
	{\Delta _2} &= \sum\limits_{m = 0}^t {\frac{{t!{e^{ - {\eta _1}}}}}{{m!\left( {t - m} \right)!}}} \int_1^\infty  {\int_{y{\Theta _{1,p}} - 1}^\infty  {{e^{ - \left( {{\eta _1} + \frac{1}{{{\delta _1}}}} \right)z}}{z^{m + {L_p} - 1}}dz} } {e^{ - \frac{{y - 1}}{{{\delta _{e1}}}}}}{\left( {y - 1} \right)^{{L_{e1}} - 1}}dy\\
	&= \sum\limits_{m = 0}^t {\sum\limits_{k = 0}^{m + {L_p} - 1} {\sum\limits_{i = 0}^k {\sum\limits_{j = 0}^i {{A_5}} } } } \int_0^\infty  {{e^{ - \left( {{\Theta _{1,p}}\left( {{\eta _1} + \frac{1}{{{\delta _1}}}} \right) + \frac{1}{{{\delta _{e1}}}}} \right)z}}{z^{j + {L_{e1}} - 1}}dz} \\
	&= \sum\limits_{m = 0}^t {\sum\limits_{k = 0}^{m + {L_p} - 1} {\sum\limits_{i = 0}^k {\sum\limits_{j = 0}^i {{A_5}} } } } \left( {j + {L_{e1}} - 1} \right)!{\left( {{\Theta _{1,p}}\left( {{\eta _1} + \frac{1}{{{\delta _1}}}} \right) + \frac{1}{{{\delta _{e1}}}}} \right)^{ - j - {L_{e1}}}}
	\label{delta_2}	
    \end{aligned}
	\end{equation}
\hrulefill
\end{figure*}
\setcounter{equation}{\value{TempEqCnt2}}
\setcounter{equation}{17} %指定跨栏公式的序号		

\newcounter{TempEqCnt3}
\setcounter{TempEqCnt3}{\value{equation}}
\setcounter{equation}{17} %指定跨栏公式的序号
\begin{figure*}[ht]
	\begin{equation}
		\begin{aligned}
			{\Delta _3} &= \sum\limits_{i = 1}^I {\frac{{\pi \sqrt {{b_1} - b_1^2} }}{{Ib_1^{{a_1}}}}} \int_1^\infty  {\frac{{{e^{ - \left( {{\eta _1} + \frac{1}{{{\delta _1}}}} \right)\frac{{y{\Theta _{1,p}} - 1}}{{{b_1}}}}}{{\left( {y{\Theta _{1,p}} - 1} \right)}^{n + {L_p}}}{y^t}{{\left( {y - 1} \right)}^{{L_{e1}} - 1}}}}{{{{\left( {{\eta _2}\left( {y{\Theta _{1,p}} - 1} \right) + {b_1}y + {b_1}{\eta _2}} \right)}^{m + {L_c} - k}}{e^{\frac{{y - 1}}{{{\delta _{e1}}}}}}}}dy} \\
			&= \sum\limits_{i = 1}^I {\sum\limits_{j = 0}^{n + {L_p}} {\sum\limits_{q = 0}^t {{A_6}} } } \int_0^\infty  {\frac{{{e^{ - \left( {\left( {{\eta _1} + \frac{1}{{{\delta _1}}}} \right)\frac{{{\Theta _{1,p}}}}{{{b_1}}} + \frac{1}{{{\delta _{e1}}}}} \right)u}}{u^{j + q + {L_{e1}} - 1}}}}{{{{\left( {\left( {{\eta _2}{\Theta _{1,p}} + {b_1}} \right)u + {b_2}} \right)}^{m + {L_c} - k}}}}dy} \\
			&= \sum\limits_{i = 1}^I {\sum\limits_{j = 0}^{n + {L_p}} {\sum\limits_{q = 0}^t {\frac{{{A_6}{b_3}^{{a_2}}G_{1,2}^{2,1}\left[ {\left( {\left( {{\eta _1} + \frac{1}{{{\delta _1}}}} \right)\frac{{{\Theta _{1,p}}}}{{{b_1}}} + \frac{1}{{{\delta _{e1}}}}} \right){b_3}\left| {_{ - {a_2},0}^{1 - j - q - {L_{e1}}}} \right.} \right]}}{{{{\left( {{\eta _2}{\Theta _{1,p}} + {b_1}} \right)}^{m + {L_c} - k}}\Gamma \left( {m + {L_c} - k} \right)}}} } }
			\label{delta_3}	
		\end{aligned}
	\end{equation}
	\hrulefill
\end{figure*}
\setcounter{equation}{\value{TempEqCnt3}}
\setcounter{equation}{18} %指定跨栏公式的序号

\textbf{{3) Scenario III}}: ${\theta _{e,\max }} \in {\Delta _{{L_{ec}} + {L_{e1}}}}$

  In this case, the common streams is eavesdropped completely while the private streams is eavesdropped partly.
Thus, ${P_{{\mathrm{scp}},1}^{\mathrm{III}}}$ is expressed as
\begin{equation}
	\begin{aligned}
						&{P_{{\mathrm{scp}},1}^{\mathrm{III}}} \\
						&= \Pr \left\{ {{{\log }_2}\left( {\frac{{1 + {\gamma _{1,c}}}}{{1 + {\gamma _{e,c}^{\mathrm{III}}}}}} \right) > {R_{1,c}^{\mathrm{th}}},{{\log }_2}\left( {\frac{{1 + {\gamma _{1,p}}}}{{1 + {\gamma _{e,p1}^{\mathrm{III}}}}}} \right) > {R_{1,p}^{\mathrm{th}}}} \right\}\\
						& = \Pr \left\{ {{X_{ec}} < \left( {\frac{{{X_{1c}}}}{{{{X}_1}{\eta _2}}} - \frac{{{\eta _1}}}{{{\eta _2}}}} \right)\left( {{{X}_3} + {\delta _{e1}}{X_{ep1}}} \right),{X_{ep1}} < } \right.\\
						& \;\quad\;\quad  \left. {\left( {\frac{{{X_1}}}{{{\Theta _{1,p}}}} - 1} \right)\frac{{{X_3}}}{{{\delta _{e1}}}},}{{X_{1c}} > {{X}_1}{\eta _1},{{X}_1} > {\Theta _{1,p}},{{X}_3} > 1} \right\}\\
						& = {{\mathbb{E}}_{{X_{1c}},{{X}_1},{{X}_3}}}\left[ {{\Delta _4}} \right],
	\end{aligned}
	\label{Psop3_medium}
\end{equation}
where
${{X}_3} = 1 + {\delta _{e2}}{X_{ep2}}$
and 
${\Delta _4} = \int_0^{\frac{{Q{{X}_3}}}{{{\delta _{e1}}}}} {{F_{{X_{ec}}}}\left( {P\left( {{{X}_3} + {\delta _{e1}}y} \right)} \right){f_{{X_{ep1}}}}\left( y \right)dy} $,
$P = \frac{{{X_{1c}}}}{{{{X}_1}{\eta _2}}} - \frac{{{\eta _1}}}{{{\eta _2}}}$,
and 
$Q = \frac{{{{X}_1}}}{{{\Theta _{1,p}}}} - 1$.
Based on (\ref{xpdf}) and (\ref{xcdf}) and utilizing \cite[(3.351.1)]{Gradshteyn2007Book}, ${\Delta _4}$ is obtained as
\begin{equation}
	\begin{aligned}
					{\Delta _4} & = 1 - {e^{ - \frac{{Q{{X}_3}}}{{{\delta _{e1}}}}}}\sum\limits_{t = 0}^{{L_{e1}} - 1} {\frac{1}{{t!}}{{\left( {\frac{{Q{{X}_3}}}{{{\delta _{e1}}}}} \right)}^t}} \\
                     &- \sum\limits_{t = 0}^{{L_{ec}} - 1} {\sum\limits_{m = 0}^t {\frac{{{B_1}{P^t}{X}_3^{t - m}{e^{ - P{{X}_3}}}}}{{{{\left( {P{\delta _{e1}} + 1} \right)}^{m + {L_{e1}}}}}}} } \\
					&+ \sum\limits_{t = 0}^{{L_{ec}} - 1} {\sum\limits_{m = 0}^t {\sum\limits_{k = 0}^{m + {L_{e1}} - 1} {\frac{{{B_1}{P^t}{{\left( {\frac{{Q{{X}_3}}}{{{\delta _{e1}}}}} \right)}^k}{e^{ - \frac{{{{X}_1}}}{{{\Theta _{1,p}}}}P{{X}_3}}}}}{{k!{X}_3^{m - t}{e^{\frac{{Q{{X}_3}}}{{{\delta _{e1}}}}}}{{\left( {P{\delta _{e1}} + 1} \right)}^{m + {L_{e1}} - k}}}}} } },
	\end{aligned}
	\label{Delta_4}
\end{equation}
where
${B_1} = \frac{{\delta _{e1}^m\left( {m + {L_{e1}} - 1} \right)!}}{{m!\left( {t - m} \right)!\Gamma \left( {{L_{e1}}} \right)}}$.
Substituting (\ref{Delta_4}) into (\ref{Psop3_medium}), ${P_{{\mathrm{scp}},1}^{\mathrm{III}}}$ is obtained as
\begin{equation}
	\begin{aligned}
		{P_{{\mathrm{scp}},1}^{\mathrm{III}}} &= {\Delta _5} - \sum\limits_{t = 0}^{{L_{e1}} - 1} {\frac{1}{{t!}}{\Delta _6}}  - \sum\limits_{t = 0}^{{L_{ec}} - 1} {\sum\limits_{m = 0}^t {{\Delta _7}} }  \\
		&+ \sum\limits_{t = 0}^{{L_{ec}} - 1} {\sum\limits_{m = 0}^t {\sum\limits_{k = 0}^{m + {L_{e1}} - 1} {{\Delta _8}} } },
	\end{aligned}
	\label{H232}
\end{equation}
where
${\Delta _5} = \int_1^\infty  {\int_{{\Theta _{1,p}}}^\infty  {\int_{y{\eta _1}}^\infty  {{f_{{X_{1c}}}}} \left( x \right)dx{f_{{{X}_1}}}\left( y \right)dy{f_{{{X}_3}}}\left( z \right)dz} } $, 
${\Delta _6} = {{\mathbb{E}}_{{X_{1c}},{{X}_1},{{X}_3}}}\left[ {{e^{ - \left( {\frac{{{{X}_1}}}{{{\Theta _{1,p}}}} - 1} \right)\frac{{{{X}_3}}}{{{\delta _{e1}}}}}}{{\left( {\left( {\frac{{{{X}_1}}}{{{\Theta _{1,p}}}} - 1} \right)\frac{{{{X}_3}}}{{{\delta _{e1}}}}} \right)}^t}} \right]$, 
${\Delta _7} = {{\mathbb{E}}_{{X_{1c}},{{X}_1},{{X}_3}}}\left[ {\frac{{{B_1}{X}_3^{t - m}{{\left( {\frac{{{X_{1c}}}}{{{{X}_1}{\eta _2}}} - \frac{{{\eta _1}}}{{{\eta _2}}}} \right)}^t}{e^{ - \left( {\frac{{{X_{1c}}}}{{{{X}_1}{\eta _2}}} - \frac{{{\eta _1}}}{{{\eta _2}}}} \right){{X}_3}}}}}{{{{\left( {\left( {\frac{{{X_{1c}}}}{{{{X}_1}{\eta _2}}} - \frac{{{\eta _1}}}{{{\eta _2}}}} \right){\delta _{e1}} + 1} \right)}^{m + {L_{e1}}}}}}} \right]$,\\
and 
${\Delta _8} = {{\mathbb{E}}_{{X_{1c}},{{X}_1},{{X}_3}}}\left[ {\frac{{{B_1}{{\left( {\left( {\frac{{{X_1}}}{{{\Theta _{1,p}}}} - 1} \right)\frac{{{X_3}}}{{{\delta _{e1}}}}} \right)}^k}{{\left( {\frac{{{X_{1c}}}}{{{X_1}{\eta _2}}} - \frac{{{\eta _1}}}{{{\eta _2}}}} \right)}^t}}}{{{{\left( {\left( {\frac{{{X_{1c}}}}{{{X_1}{\eta _2}}} - \frac{{{\eta _1}}}{{{\eta _2}}}} \right){\delta _{e1}} + 1} \right)}^{m + {L_{e1}} - k}}}}} \right.$ $\times\left. {\frac{{{e^{ - \frac{{{X_1}}}{{{\Theta _{1,p}}}}\left( {\frac{{{X_{1c}}}}{{{X_1}{\eta _2}}} - \frac{{{\eta _1}}}{{{\eta _2}}}} \right){X_3}}}}}{{k!X_3^{m - t}{e^{\left( {\frac{{{X_1}}}{{{\Theta _{1,p}}}} - 1} \right)\frac{{{X_3}}}{{{\delta _{e1}}}}}}}}} \right]$.

Utilizing \cite[(3.351.2)]{Gradshteyn2007Book}, ${\Delta _5}$ is obtained as
\begin{equation}
	\begin{aligned}
				{\Delta _5} &= \int_1^\infty  {\int_{{\Theta _{1,p}}}^\infty  {{{\bar F}_{{X_{1c}}}}\left( {y{\eta _1}} \right){f_{{{X}_1}}}\left( y \right)dy{f_{{{X}_3}}}\left( z \right)dz} } \\
				&= \sum\limits_{t = 0}^{{L_c} - 1} {\frac{{\eta _1^t}}{{t!}}} \int_{{\Theta _{1,p}}}^\infty  {\frac{{{e^{ - y{\eta _1}}}{y^t}{e^{ - \frac{{y - 1}}{{{\delta _1}}}}}{{\left( {y - 1} \right)}^{{L_p} - 1}}}}{{\delta _1^{{L_p}}\left( {{L_p} - 1} \right)!}}dy} \\
				&= \sum\limits_{t = 0}^{{L_c} - 1} {\sum\limits_{m = 0}^t {\frac{{\eta _1^t{e^{ - {\eta _1}}}\Gamma \left( {m + {L_p},\left( {{\eta _1} + \frac{1}{{{\delta _1}}}} \right)\left( {{\Theta _{1,p}} - 1} \right)} \right)}}{{\delta _1^{{L_p}}m!\left( {t - m} \right)!\left( {{L_p} - 1} \right)!{{\left( {{\eta _1} + \frac{1}{{{\delta _1}}}} \right)}^{m + {L_p}}}}}} }.
	\end{aligned}
\end{equation}
Based on  (\ref{xcdf}) and (\ref{X_PDF}) and utilizing \cite[(3.351.3), (7.811.5)]{Gradshteyn2007Book}, ${\Delta _6}$ is obtained as
\begin{equation}
	\begin{aligned}
				{\Delta _6} &= \sum\limits_{m = 0}^{{L_c} - 1} {\sum\limits_{n = 0}^m {\sum\limits_{i = 0}^{{L_p} - 1} {{B_2}} } }\int_1^\infty  {\int_0^\infty  {{e^{ - \left( {z + \frac{{{\Theta _{1,p}}{\delta _{e1}}}}{{{\delta _1}}} + {\Theta _{1,p}}{\delta _{e1}}{\eta _1}} \right)u}}} }   \\
				& \times {u^{t + n + i}}du{z^t}\frac{{{e^{ - \frac{{z - 1}}{{{\delta _{e2}}}}}}{{\left( {z - 1} \right)}^{{L_{e2}} - 1}}}}{{\delta _{e2}^{{L_{e2}}}\left( {{L_{e2}} - 1} \right)!}}dz\\
				& = \sum\limits_{m = 0}^{{L_c} - 1} {\sum\limits_{n = 0}^m {\sum\limits_{i = 0}^{{L_p} - 1} {\sum\limits_{j = 0}^t {\frac{{{B_2}\left( {t + n + i} \right)!t!{\Xi _1}}}{{j!\left( {t - j} \right)!\delta _{e2}^{{L_{e2}}}\left( {{L_{e2}} - 1} \right)!}}} } } },
	\end{aligned}
    \label{delta_5_dayu}
\end{equation}
where
${B_2} = \frac{{{{\left( {{\Theta _{1,p}} - 1} \right)}^{{L_p} - 1 - i}}\Theta _{1,p}^{m + 1 + i}\delta _{e1}^{i + n + 1}\eta _1^m}}{{i!\left( {{L_p} - 1 - i} \right)!n!\left( {m - n} \right)!\delta _1^{{L_p}}{e^{{\eta _1}{\Theta _{1,p}} + \frac{{{\Theta _{1,p}} - 1}}{{{\delta _1}}}}}}}$, 
${\Xi _1} =\frac{{{{\left( {1 + {\eta _3}} \right)}^{{a_4} - {a_3}}}}}{{\Gamma \left( {{a_3}} \right)}}G_{1,2}^{2,1}\left[ {\frac{{\left( {1 + {\eta _3}} \right)}}{{{\delta _{e2}}}}\left| {_{{a_3} - {a_4},0}^{1 - {a_4}}} \right.} \right]$, 
${a_3} = t + n + i + 1$,
${a_4} = j + {L_{e2}}$,
and
${\eta _3} = \frac{{{\Theta _{1,p}}{\delta _{e1}}}}{{{\delta _1}}} + {\Theta _{1,p}}{\delta _{e1}}{\eta _1}$.
Based on  (\ref{xpdf}) and (\ref{X_PDF}), ${\Delta _7}$ is obtained as
\begin{equation}
	\begin{aligned}
			{\Delta _7} &= {B_1}{{\mathbb{E}}_{{{X}_1},{{X}_3}}}\left[ {\frac{{X_3^{t - m}}}{{\Gamma \left( {{L_c}} \right)}}\int_{{X_1}{\eta _1}}^\infty  {\frac{{{{\left( {\frac{x}{{{X_1}{\eta _2}}} - \frac{{{\eta _1}}}{{{\eta _2}}}} \right)}^t}{e^{ - x}}{x^{{L_c} - 1}}}}{{{{\left( {\left( {\frac{x}{{{X_1}{\eta _2}}} - \frac{{{\eta _1}}}{{{\eta _2}}}} \right){\delta _{e1}} + 1} \right)}^{m + {L_{e1}}}}}}} } \right.\\
			&\times\left. {{e^{ - \left( {\frac{x}{{{X_1}{\eta _2}}} - \frac{{{\eta _1}}}{{{\eta _2}}}} \right){X_3}}}dx} \right]\\
			&= \sum\limits_{n = 0}^{{L_c} - 1} {{B_3}} {{\mathbb{E}}_{{{X}_1},{{X}_3}}}{\left[ {\frac{{X_3^{t - m}X_1^{{L_c}}{e^{ - {X_1}{\eta _1}}}}}{{\Gamma \left( {{L_c}} \right)}}} \right.}\\
			&\times{\left. {\int_0^\infty  {\frac{{{e^{ - \left( {{X_3} + {X_1}{\eta _2}} \right)u}}{u^{n + t}}}}{{{{\left( {u{\delta _{e1}} + 1} \right)}^{m + {L_{e1}}}}}}du} } \right]}\\
			&= \sum\limits_{n = 0}^{{L_c} - 1} {\frac{{{B_{\mathrm{3}}}\delta _1^{ - {L_p}}\delta _{e2}^{ - {L_{e2}}}{\Xi _{\mathrm{2}}}}}{{\Gamma \left( {{L_c}} \right)\left( {{L_p} - 1} \right)!\left( {{L_{e2}} - 1} \right)!}}},
	\end{aligned}
	\label{Delta_6}
\end{equation}
where
${B_3} = \frac{{{B_1}\left( {{L_c} - 1} \right)!\eta _1^{{L_c} - 1 - n}\eta _2^{n + 1}}}{{n!\left( {{L_c} - 1 - n} \right)!}}$
and
${\Xi _{\mathrm{2}}} = \int_1^\infty  {\frac{{{\nabla _1}{z^{t - m}}{{\left( {z - 1} \right)}^{{L_{e2}} - 1}}}}{{{e^{\frac{{z - 1}}{{{\delta _{e2}}}}}}}}} dz$, \\
${\nabla _1} = \int_{{\Theta _{1,p}}}^\infty  {\int_0^\infty  {\frac{{{e^{ - \left( {z + y{\eta _2}} \right)u}}{u^{n + t}}}}{{{{\left( {u{\delta _{e1}} + 1} \right)}^{m + {L_{e1}}}}}}du} } \frac{{{y^{{L_c}}}{{\left( {y - 1} \right)}^{{L_p} - 1}}}}{{{e^{y{\eta _1}}}{e^{\frac{{y - 1}}{{{\delta _1}}}}}}}dy$.
Then, utilizing \cite[(7.811.5)]{Gradshteyn2007Book} and (25.4.39), ${\nabla _1}$ is obtained as
\begin{equation}
	\begin{aligned}
			{\nabla _1} &= \sum\limits_{k = 0}^{{L_c}} {\frac{{\left( {{L_c}} \right)!\delta _{e1}^{ - {a_5}}{e^{ - {\eta _1}}}}}{{k!\left( {{L_c} - k} \right)!\Gamma \left( {m + {L_{e1}}} \right)}}}\\
			&\times \int_{{\Theta _{1,p}} - 1}^\infty  {G_{1,2}^{2,1}\left[ {\frac{{z + \lambda {\eta _2} + {\eta _2}}}{{{\delta _{e1}}}}\left| {_{ - {a_5},0}^{ - n - t}} \right.} \right]} \frac{{{\lambda ^{k + {L_p} - 1}}}}{{{e^{\left( {{\eta _1} + \frac{1}{{{\delta _1}}}} \right)\lambda }}}}d\lambda \\
			&= \sum\limits_{k = 0}^{{L_c}} {\frac{{\left( {{L_c}} \right)!\delta _{e1}^{ - {a_5}}{e^{ - {\eta _1}}}}}{{k!\left( {{L_c} - k} \right)!\Gamma \left( {m + {L_{e1}}} \right)}}} \\
			&\times\int_0^{\frac{1}{{{\Theta _{1,p}} - 1}}} {G_{1,2}^{2,1}\left[ {\frac{{\left( {z + {\eta _2}} \right)u + {\eta _2}}}{{{\delta _{e1}}u}}\left| {_{ - {a_5},0}^{ - n - t}} \right.} \right]} \frac{{{e^{ - \left( {{\eta _1} + \frac{1}{{{\delta _1}}}} \right)\frac{1}{u}}}}}{{{u^{k + {L_p} + 1}}}}du\\
			&= \sum\limits_{k = 0}^{{L_c}} {\sum\limits_{i = 1}^N {{B_4}} } G_{1,2}^{2,1}\left[ {\frac{{{b_4}z + \left( {{b_4} + 1} \right){\eta _2}}}{{{\delta _{e1}}{b_4}}}\left| {_{ - {a_5},0}^{ - n - t}} \right.} \right],
	\end{aligned}
	\label{nabla_1}
\end{equation}
where
${a_5} = n + t + 1 - m - {L_{e1}}$,
${B_4} = \frac{{\pi \left( {{L_c}} \right)!\delta _{e1}^{ - {a_5}}{e^{ - {\eta _1}}}\sqrt {\frac{1}{{{\Theta _{1,p}} - 1}} - {b_4}} }}{{Nk!\left( {{L_c} - k} \right)!\Gamma \left( {m + {L_{e1}}} \right){e^{\left( {{\eta _1} + \frac{1}{{{\delta _1}}}} \right)\frac{1}{{{b_4}}}}}{{\left( {{b_4}} \right)}^{k + {L_p} + \frac{1}{2}}}}}$,
${b_4} = \frac{{1 + \cos \frac{{\left( {2i - 1} \right)\pi }}{{2N}}}}{{2\left( {{\Theta _{1,p}} - 1} \right)}}$
and
$N$ is the summation items that reflect accuracy versus complexity.
Based on (\ref{nabla_1}) and utilizing (25.4.45), ${\Xi _{\mathrm{2}}}$ is obtained as (\ref{Xi_2}), shown at the top of next page,
where 
$V$ is the summation items that reflect accuracy versus complexity,
${t_u}$ is the $u$th zero of Leguerre polynomials
and
${w_u}$ is Gaussian weight, which are given in Table (25.9) of \cite{Abramowitz1972Book}.
With the same method, 
${\Delta _8}$ is obtained as
\newcounter{TempEqCnt4}
\setcounter{TempEqCnt4}{\value{equation}}
\setcounter{equation}{25} %指定跨栏公式的序号
\begin{figure*}[ht]
	\begin{equation}
		\begin{aligned}
			{\Xi _{\mathrm{2}}} %&= \int_1^\infty  {\frac{{{\nabla _1}{z^{t - m}}{{\left( {z - 1} \right)}^{{L_{e2}} - 1}}}}{{{e^{\frac{{z - 1}}{{{\delta _{e2}}}}}}}}} dz\\
			&= \sum\limits_{k = 0}^{{L_c}} {\sum\limits_{i = 1}^N {\sum\limits_{j = 0}^{t - m} {\frac{{\left( {t - m} \right)!{B_4}}}{{j!\left( {t - m - j!} \right)}}} } } \int_0^\infty  {G_{1,2}^{2,1}\left[ {\frac{{{b_4}f + {b_4} + \left( {{b_4} + 1} \right){\eta _2}}}{{{\delta _{e1}}{b_4}}}\left| {_{ - {a_5},0}^{ - n - t}} \right.} \right]{e^{ - \frac{f}{{{\delta _{e2}}}}}}{f^{j + {L_{e2}} - 1}}} df\\
			&= \sum\limits_{k = 0}^{{L_c}} {\sum\limits_{i = 1}^N {\sum\limits_{j = 0}^{t - m} {\frac{{\left( {t - m} \right)!{B_4}\delta _{e2}^{j + {L_{e2}}}}}{{j!\left( {t - m - j!} \right)}}} } } \sum\limits_{u = 1}^V {{w_u}} G_{1,2}^{2,1}\left[ {\frac{{{b_4}{\delta _{e2}}{t_u} + {b_4} + \left( {{b_4} + 1} \right){\eta _2}}}{{{\delta _{e1}}{b_4}}}\left| {_{ - {a_5},0}^{ - n - t}} \right.} \right]t_u^{j + {L_{e2}} - 1}\\
			&= \sum\limits_{k = 0}^{{L_c}} {\sum\limits_{i = 1}^N {\sum\limits_{j = 0}^{t - m} {\sum\limits_{u = 1}^V {\frac{{\left( {t - m} \right)!{B_4}\delta _{e2}^{j + {L_{e2}}}{w_u}}}{{j!\left( {t - m - j!} \right)}}t_u^{j + {L_{e2}} - 1}G_{1,2}^{2,1}\left[ {\frac{{{b_4}{\delta _{e2}}{t_u} + {b_4} + \left( {{b_4} + 1} \right){\eta _2}}}{{{\delta _{e1}}{b_4}}}\left| {_{ - {a_5},0}^{ - n - t}} \right.} \right]} } } }
			\label{Xi_2}	
		\end{aligned}
	\end{equation}
	\hrulefill
\end{figure*}
\setcounter{equation}{\value{TempEqCnt4}}
\setcounter{equation}{26} %指定跨栏公式的序号

\begin{equation}
	\begin{aligned}
					{\Delta _8} &=\sum\limits_{n = 0}^{{L_c} - 1} {\frac{{{B_3}\delta _1^{ - {L_p}}\delta _{e2}^{ - {L_{e2}}}{\Xi _{\mathrm{3}}}}}{{k!\Gamma \left( {{L_c}} \right)\left( {{L_p} - 1} \right)!\left( {{L_{e2}} - 1} \right)!}}},
	\end{aligned}
	\label{delta_4_dayu_1}
\end{equation}
where
${\Xi _{\mathrm{3}}} = \int_1^\infty  {\frac{{{\nabla _2}{{\left( {z - 1} \right)}^{{L_{e2}} - 1}}}}{{{z^{m - t}}{e^{\frac{{z - 1}}{{{\delta _{e2}}}}}}}}dz} $,
${\nabla _2} = \int_{{\Theta _{1,p}}}^\infty  {\int_0^\infty  {\frac{{{e^{ - \left( {\frac{y}{{{\Theta _{1,p}}}}z + y{\eta _2}} \right)u}}{u^{n + t}}}}{{{{\left( {u{\delta _{e1}} + 1} \right)}^{m + {L_{e1}} - k}}}}du} \frac{{{e^{ - \left( {\frac{y}{{{\Theta _{1,p}}}} - 1} \right)\frac{z}{{{\delta _{e1}}}}}}{y^{{L_c}}}{{\left( {y - 1} \right)}^{{L_p} - 1}}}}{{{{\left( {\left( {\frac{y}{{{\Theta _{1,p}}}} - 1} \right)\frac{z}{{{\delta _{e1}}}}} \right)}^{ - k}}{e^{y{\eta _1}}}{e^{\frac{{y - 1}}{{{\delta _1}}}}}}}dy} $.
By utilizing \cite[(7.811.5)]{Gradshteyn2007Book} and (25.4.45), ${\nabla _2}$ is obtained as as (\ref{nabla_2}), shown at the top of next page,
where
${B_5} = \frac{{\left( {{L_c}} \right)!\left( {{L_p} - 1} \right)!\Theta _{1,p}^{{L_c} + j + 1}{{\left( {{\Theta _{1,p}} - 1} \right)}^{{L_p} - 1 - j}}{e^{ - \left( {{\Theta _{1,p}}{\eta _1} + \frac{{{\Theta _{1,p}} - 1}}{{{\delta _1}}}} \right)}}}}{{i!\left( {{L_c} - i} \right)!j!\left( {{L_p} - 1 - j} \right)!\delta _{e1}^{k + n + t + 1}\Gamma \left( {m + {L_{e1}} - k} \right)}}$, 
${a_6} = m + {L_{e1}} - k - n - t - 1$,
${b_5} = {\delta _{e1}}{\Theta _{1,p}}{\eta _1} + {\delta _{e1}}\frac{{{\Theta _{1,p}}}}{{{\delta _1}}}$, 
$K$ is the summation items that reflect accuracy versus complexity,
${{u_v}}$ is the $v$th zero of Leguerre polynomials
and
${{q_v}}$ is Gaussian weight, which are given in Table (25.9) of \cite{Abramowitz1972Book}.
Based on (\ref{nabla_2}) and utilizing (25.4.45), $\Xi _3$ is obtained as
\newcounter{TempEqCnt5}
\setcounter{TempEqCnt5}{\value{equation}}
\setcounter{equation}{27} %指定跨栏公式的序号
\begin{figure*}[ht]
	\begin{equation}
		\begin{aligned}
			{\nabla _2} &= \sum\limits_{i = 0}^{{L_c}} {\sum\limits_{j = 0}^{{L_p} - 1} {{B_5}} } {z^k}\int_0^\infty  {G_{1,2}^{2,1}\left[ {\left( {z + {\Theta _{1,p}}{\eta _2}} \right)\frac{{\left( {\lambda  + 1} \right)}}{{{\delta _{e1}}}}\left| {_{{a_6},0}^{ - n - t}} \right.} \right]} \frac{{{\lambda ^{i + j + k}}}}{{{e^{\left( {\frac{z}{{{\delta _{e1}}}} + {\Theta _{1,p}}{\eta _1} + \frac{{{\Theta _{1,p}}}}{{{\delta _1}}}} \right)\lambda }}}}d\lambda \\
			&= \sum\limits_{i = 0}^{{L_c}} {\sum\limits_{j = 0}^{{L_p} - 1} {\sum\limits_{v = 1}^K {\frac{{{q_v}{B_5}\delta _{e1}^{i + j + k + 1}{z^k}}}{{{{\left( {z + {b_5}} \right)}^{i + j + k + 1}}}}} } } u_v^{i + j + k}G_{1,2}^{2,1}\left[ {\left( {z + {\Theta _{1,p}}{\eta _2}} \right)\left( {\frac{{{u_v}}}{{z + {b_5}}} + \frac{1}{{{\delta _{e1}}}}} \right)\left| {_{{a_6},0}^{ - n - t}} \right.} \right]
			\label{nabla_2}	
		\end{aligned}
	\end{equation}
	\hrulefill
\end{figure*}
\setcounter{equation}{\value{TempEqCnt5}}
\setcounter{equation}{28} %指定跨栏公式的序号

\begin{equation}
	\begin{aligned}
			{\Xi _{\mathrm{3}}}	&= \sum\limits_{i = 0}^{{L_c}} {\sum\limits_{j = 0}^{{L_p} - 1} {\sum\limits_{v = 1}^K {\sum\limits_{q = 0}^{k + t - m} {\sum\limits_{u = 1}^D {B_6}{\varepsilon _u} } } } }\\
			&\times{\frac{{G_{1,2}^{2,1}\left[ {\left( {\frac{{{s_u}{\delta _{e2}}}}{{{\delta _{e1}}}} + \frac{{{u_v}\left( {{\Theta _{1,p}}{\eta _2} - {b_5}} \right)}}{{{\delta _{e2}}{s_u} + 1 + {b_5}}} + {b_6}} \right)\left| {_{{a_6},0}^{ - n - t}} \right.} \right]}}{{{{\left( {{\delta _{e2}}{s_u} + 1 + {b_5}} \right)}^{i + j + k + 1}}\delta _{e2}^{ - q - {L_{e2}}}s_u^{1 - q - {L_{e2}}}}}},
	\end{aligned}
	\label{nabla_3}
\end{equation}
where
${B_6} = \frac{{\left( {k + t - m} \right)!{q_v}{B_5}\delta _{e1}^{i + j + k + 1}u_v^{i + j + k}}}{{q!\left( {k + t - m - q} \right)}}$,
${b_6} = {u_v} + \frac{{1 + {\Theta _{1,p}}{\eta _2}}}{{{\delta _{e1}}}}$,
$D$ is the summation items that reflect accuracy versus complexity,
${s_u}$ is the $u$th zero of Leguerre polynomials
and
${\varepsilon _u}$ is Gaussian weight, which are given in Table (25.9) of \cite{Abramowitz1972Book}.

\textbf{4) Scenario IV}: ${\theta _{e,\max }} \in {\Delta _{{L_{ec}}}}$

In this case, $E$ only eavesdrop on common streams. 
It should noted that ${{\gamma _{1,c}}}$ and $\gamma _{e,c}^{\mathrm{IV}}$ are independent of each other in this case.
Based on (\ref{xpdf}) and (\ref{xcdf}) and utilizing \cite[(3.351.3)]{Gradshteyn2007Book}, the CDF of ${\gamma _{1,c}}$ and PDF of ${\gamma _{e,c}^{\mathrm{IV}}}$  are obtained as
\begin{equation}
	\begin{aligned}
			{F_{{\gamma _{1,c}}}}\left( x \right) &= 1 - \sum\limits_{t = 0}^{{L_c} - 1} {\sum\limits_{m = 0}^t {{\varpi _{1,c}}{x^t}{e^{ - \frac{x}{{{\delta _{1,c}}}}}}{{\left( {\frac{{{\delta _1}}}{{{\delta _{1,c}}}}x + 1} \right)}^{ - m - {L_p}}}} },
	\end{aligned}
	\label{cdf_chi}
\end{equation}
\begin{equation}
	\small
	\begin{aligned}
		&{f_{{\gamma _{e,c}}}}\left( y \right) =\frac{1}{{{\delta _{ec}}}}\sum\limits_{s = 0}^{{L_{ec}} - 1} {\sum\limits_{n = 0}^s {\left( {s + 1 - n} \right){\varpi _{e,c}}{y^s}{e^{ - \frac{y}{{{\delta _{ec}}}}}}{{\left( {\frac{{{\delta _{e2}}}}{{{\delta _{ec}}}}y + 1} \right)}^{ - n - {L_{e2}}}}} } \\
		&+ \frac{{{\delta _{e2}}}}{{{\delta _{ec}}}}\sum\limits_{s = 0}^{{L_{ec}} - 1} {\sum\limits_{n = 0}^s {\left( {s + 1 - n} \right)\left( {n + {L_{e2}}} \right){\varpi _{e,c}}{y^s}{e^{ - \frac{y}{{{\delta _{ec}}}}}}{{\left( {\frac{{{\delta _{e2}}}}{{{\delta _{ec}}}}y + 1} \right)}^{ - n - {L_{e2}} - {\mathrm{1}}}}} }  \\
		&- \sum\limits_{s = 0}^{{L_{ec}} - 2} {\sum\limits_{n = 0}^{s + 1} {\frac{1}{{{\delta _{ec}}}}{\varpi _{e,c}}\left( {s + 1} \right){y^s}{e^{ - \frac{y}{{{\delta _{ec}}}}}}{{\left( {\frac{{{\delta _{e2}}}}{{{\delta _{ec}}}}y + 1} \right)}^{ - n - {L_{e2}}}}} },
	\end{aligned}
	\label{pdf_chi}
\end{equation}
respectively, 
where
${\varpi _{1,c}} = \frac{{{\delta _1}^m\left( {m + {L_p} - 1} \right)!}}{{m!\delta _{1,c}^t\left( {{L_p} - 1} \right)!\left( {t - m} \right)!}}$
and 
${\varpi _{e,c}} = \frac{{\delta _{e2}^n\left( {n + {L_{e2}} - 1} \right)!}}{{n!\delta _{ec}^s\left( {{L_{e2}} - 1} \right)!\left( {s + 1 - n} \right)!}}$.

To facilitate the following analysis, we define
\begin{equation}
	\begin{aligned}
		\phi \left( {{c_1},{c_2},{c_3},{c_4},{c_5},{c_6}} \right) = \int_0^\infty  {\frac{{{e^{ - {c_1}y}}{y^{{c_2}}}}}{{{{\left( {{c_3}y + 1} \right)}^{{c_4}}}{{\left( {{c_5}y + 1} \right)}^{{c_6}}}}}dy} .
		\label{g1}
	\end{aligned}
\end{equation}
By utilizing \cite[(10), (11)]{Adamchik1990}, \cite[(9.31.5)]{Gradshteyn2007Book}, and \cite[(1.2)]{MathaiAM2010} in turn, we obtain
\begin{equation}
	\begin{aligned}
		&\phi \left( {{c_1},{c_2},{c_3},{c_4},{c_5},{c_6}} \right) 	= \frac{{{{\left( {{c_1}} \right)}^{ - {c_2}}}}}{{\Gamma \left( {{c_4}} \right)\Gamma \left( {{c_6}} \right)}}\\
		&\times\int_0^\infty  {G_{0,1}^{1,0}\left[ {{c_1}y\left| {_{{c_2}}^ - } \right.} \right]G_{1,1}^{1,1}\left[ {{c_3}y\left| {_0^{1 - {c_4}}} \right.} \right]G_{1,1}^{1,1}\left[ {{c_5}y\left| {_0^{1 - {c_6}}} \right.} \right]} dy\\
		&= \frac{{{{\left( {{c_1}} \right)}^{ - {c_2}}}}}{{\Gamma \left( {{c_4}} \right)\Gamma \left( {{c_6}} \right){c_1}}}\\
		&\times H_{1,0:1,1:1,1}^{1,0:1,1:1,1}\left( {_ - ^{{c_2} + 1}\left| {_0^{1 - {c_4}}\left| {_0^{1 - {c_6}}\left| {\frac{{{c_3}}}{{{c_1}}},\frac{{{c_5}}}{{{c_1}}}} \right.} \right.} \right.} \right),
	\end{aligned}
\end{equation}
where 
${H_{c,d:p,r:\alpha  ,\beta }^{b,0:m,n:\gamma ,\varepsilon }\left[  \cdot  \right]}$ is the extended generalized bivariate Fox's H-function as defined by \cite[(2.57)]{MathaiAM2010}.

The SCP of $U_1$ in this scenario, ${P_{{\mathrm{scp}},1}^{\mathrm{IV}}}$ is obtained as
\begin{equation}
	\begin{aligned}
			{P_{{\mathrm{scp}},1}^{\mathrm{IV}}} &= \Pr \left\{ {\frac{{1 + {\gamma _{1,c}}}}{{1 + {\gamma _{e,c}^{\mathrm{IV}}}}} > {\Theta _{1,c}}} \right\}\\
			&= \int_0^\infty  {{{\bar F}_{{\gamma _{1,c}}}}\left( {{\Theta _{1,c}}y + {\Theta _{1,c}} - 1} \right)} {f_{{\gamma _{e,c}^{\mathrm{IV}}}}}\left( y \right)dy\\
			&= \sum\limits_{t = 0}^{{L_c} - 1} {\sum\limits_{m = 0}^t {{\varpi _{1,c}}} } \sum\limits_{s = 0}^{{L_{ec}} - 1} {\sum\limits_{n = 0}^s {{\varpi _{e,c}}} } \\
			&\times\left( {s + 1 - n} \right)\left( {\frac{{{\Delta _9}}}{{{\delta _{ec}}}} + \frac{{{\delta _{e2}}}}{{{\delta _{ec}}}}\left( {n + {L_{e2}}} \right){\Delta _{10}}} \right)\\
				&- \sum\limits_{t = 0}^{{L_c} - 1} {\sum\limits_{m = 0}^t {{\varpi _{1,c}}} } \sum\limits_{s = 0}^{{L_{ec}} - 2} {\sum\limits_{n = 0}^{s + 1} {\frac{{{\varpi _{e,c}}}}{{{\delta _{ec}}}}\left( {s + 1} \right)} } {\Delta _9},
	\end{aligned}
	\label{pdf_chi}
\end{equation}
where 
\begin{equation}
	\begin{aligned}
			{\Delta _9} 
			&= {e^{ - \frac{{{\Theta _{1,c}} - 1}}{{{\delta _{1,c}}}}}}\sum\limits_{o = 0}^t {\frac{{\Theta _{1,c}^ot!}}{{o!\left( {t - o} \right)!}}} {\left( {{\Theta _{1,c}} - 1} \right)^{t - o}}\\
			&\times{\left( {\frac{{{\delta _1}}}{{{\delta _{1,c}}}}\left( {{\Theta _{1,c}} - 1} \right) + 1} \right)^{ - m - {L_p}}} \phi \left( {\frac{{{\Theta _{1,c}}}}{{{\delta _{1,c}}}} + \frac{1}{{{\delta _{ec}}}},s + o,} \right.\\
			&\left. {\frac{{{\delta _1}{\Theta _{1,c}}}}{{{\delta _1}\left( {{\Theta _{1,c}} - 1} \right) + {\delta _{1,c}}}},m + {L_p},\frac{{{\delta _{e2}}}}{{{\delta _{ec}}}},n + {L_{e2}}} \right),
	\end{aligned}
\end{equation}
and 
\begin{equation}
	\begin{aligned}
			{\Delta _{10}} &= {e^{ - \frac{{{\Theta _{1,c}} - 1}}{{{\delta _{1,c}}}}}}\sum\limits_{o = 0}^t {\frac{{\Theta _{1,c}^ot!}}{{o!\left( {t - o} \right)!}}} {\left( {{\Theta _{1,c}} - 1} \right)^{t - o}}\\
			&\times{\left( {\frac{{{\delta _1}}}{{{\delta _{1,c}}}}\left( {{\Theta _{1,c}} - 1} \right) + 1} \right)^{ - m - {L_p}}} \phi \left( {\frac{{{\Theta _{1,c}}}}{{{\delta _{1,c}}}} + \frac{1}{{{\delta _{ec}}}},s + o,} \right.\\
			&\left. {\frac{{{\delta _1}{\Theta _{1,c}}}}{{{\delta _1}\left( {{\Theta _{1,c}} - 1} \right) + {\delta _{1,c}}}},m + {L_p},\frac{{{\delta _{e2}}}}{{{\delta _{ec}}}},n + {L_{e2}} + 1} \right).
	\end{aligned}
\end{equation}

The analytical expressions provided in this section are complicated since many factors affect the secrecy performance of $U_1$, specifically, 
the transmit power, power allocation coefficient, the target data rate, and relative locations between all the legitimate users and illegitimate receivers.

\section{Numerical Results}
\label{sec:NumericalResults}

This section presents simulation and numerical results to verify the secrecy outage performance of mmWave RSMA systems with the considered beamforming scheme. The noise power is  set at  ${\sigma ^2} =  - 71$ dBm, and the path-loss model is set as $\alpha  = 4.14$  \cite{WangC2016TWC, JuY2019TWC, RappaportST2017TCAP}.
In all the figures, `Sim' and `Ana' denote the simulation and numerical results, respectively.
\begin{figure}[t]
	\centering
	\subfigure[${P_{{\mathrm{sop}},1}^{\mathrm{I}}}$ for varying $r_1$ with ${L_{e1}} = 2$. ]{
		\label{fig2a}
		\includegraphics[width = 0.231 \textwidth]{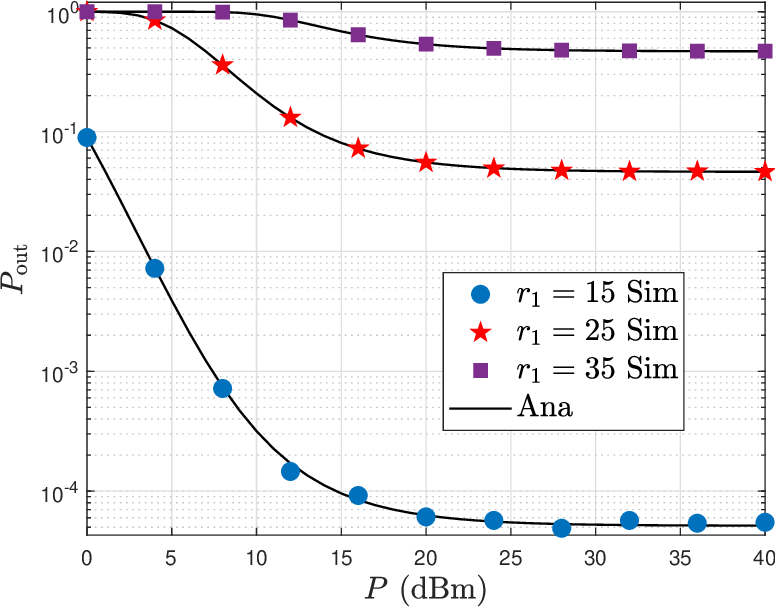}}
	\subfigure[${P_{{\mathrm{sop}},1}^{\mathrm{II}}}$ for varying $r_1$ with ${L_{ec}} = 2$ and ${L_{e1}} = L_p$. ]{
		\label{fig2b}
		\includegraphics[width = 0.231 \textwidth]{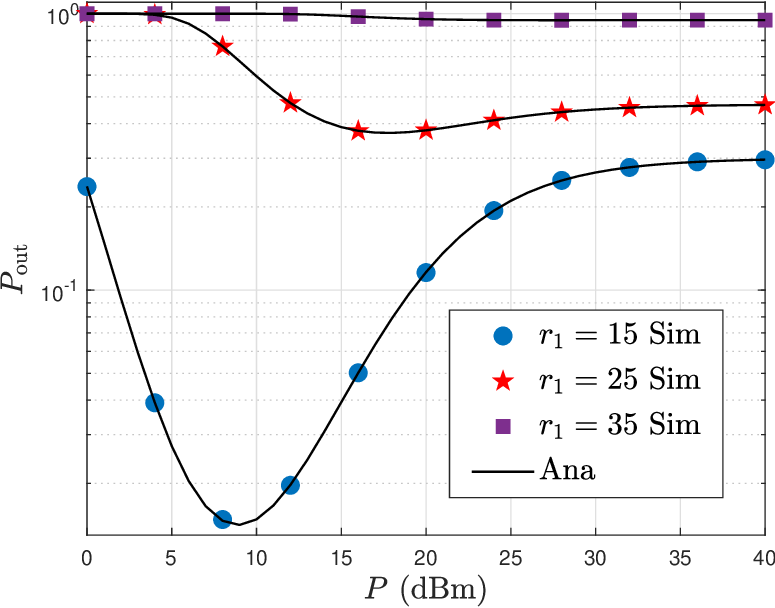}}
	\subfigure[${P_{{\mathrm{sop}},1}^{\mathrm{III}}}$ for varying $r_1$ with ${L_{e1}} = 1$, ${L_{ec}} = L_c$ and ${L_{e2}} = L - {L_{ec}} - {L_{e1}}$. ]{
		\label{fig2c}
		\includegraphics[width = 0.231 \textwidth]{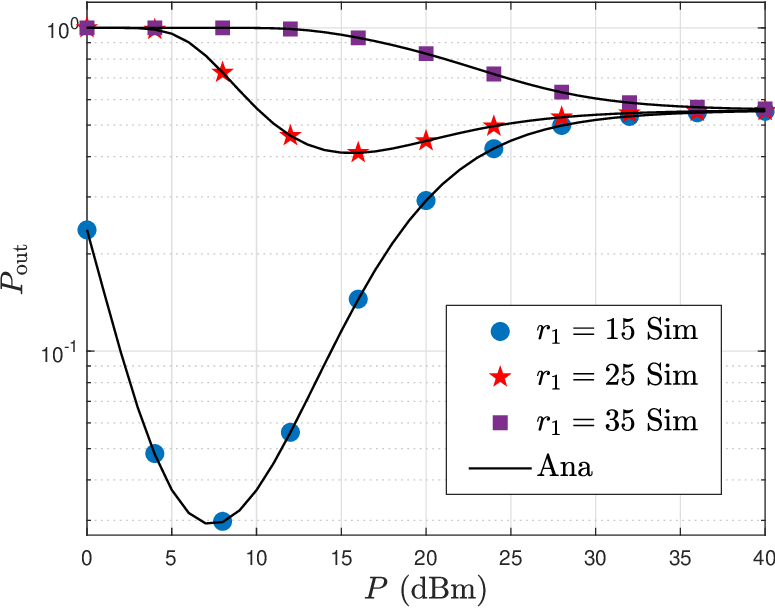}}
	\subfigure[${P_{{\mathrm{sop}},1}^{\mathrm{IV}}}$ for varying $r_1$ with ${L_{ec}} = 2$ and ${L_{e2}} = L - {L_{ec}}$. ]{
		\label{fig2d}
		\includegraphics[width = 0.231 \textwidth]{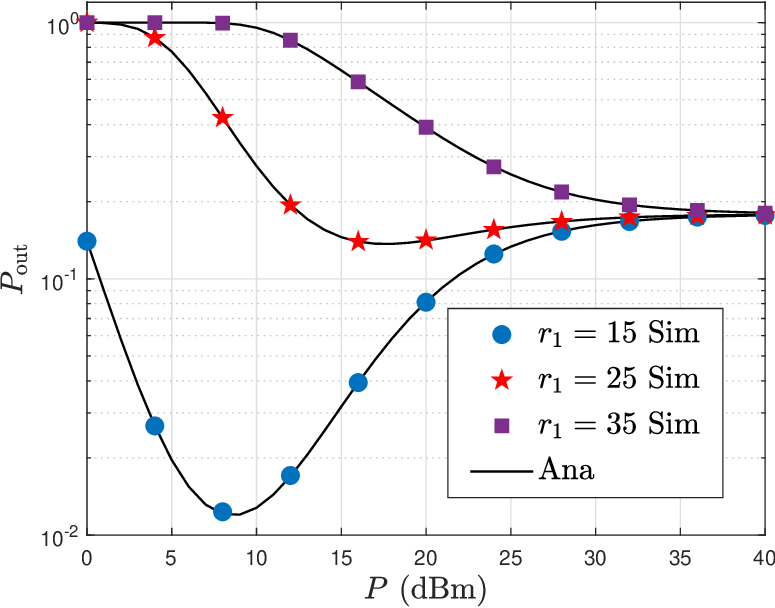}}
	\caption{SOP versus the $P$ with $\tau_c =\tau_1 =\tau_2 = \frac{1}{3}$, $r_e = 30$, $L = 8$, $L_c = 4$, and 
		$R_{{1,c}}^{ {\mathrm{th}}} = R_{1,p}^{\mathrm{th}} = 0.1$.}
	\label{fig2}
\end{figure}

Fig. \ref{fig2} demonstrates the impact of $P$ for varying $r_1$ on SOP. 
In Fig. \ref{fig2a}, one can easily observe that the secrecy outage performance of the mmWave RSMA systems is enhanced while increasing $P$. There is a floor for the SOP, which is independent of $P$, which has been testified in \cite{LeiH2017CL} and stated in \textit{Remark 1}. 
Furthermore, the SOP with lower $r_1$ outperforms that with larger $r_1$ since lower $r_1$ denotes weak path loss on $U_1$. 
One interesting result is found from Fig.s \ref{fig2b} - \ref{fig2d} that the secrecy outage performance of $U_1$ initially decreases as $P$ increases and then increases to a constant. The reason is given as follows.
In Fig. \ref{fig2b}, part of the common stream and all the private streams are wiretapped, and decoding the common stream is the bottleneck in the scenarios with lower $r_1$ and 
decoding the private stream is the bottleneck in the scenarios with larger $r_1$. 
However, in Fig. \ref{fig2c}, all of the common stream and the private stream are wiretapped, and decoding the private stream is the bottleneck in all the scenarios.
In Fig. \ref{fig2d}, only part of the common stream is wiretapped. 
Moreover, it can be found that the SOP floor is independent of $r_1$ in the scenarios where the bottleneck is decoding the common stream. 
Based on all the subfigures in Fig. \ref{fig2}, we also find that the effect from the location of $E$ on the difference between the SOP of $U_1$ with different $r_1$ is weakening. 

\begin{figure}[t]
	\centering
	\subfigure[${P_{{\mathrm{sop}},1}^{\mathrm{I}}}$ for varying $L_c$ with ${L_{e1}} = 2$.  ]{
		\label{fig3a}
		\includegraphics[width = 0.231 \textwidth]{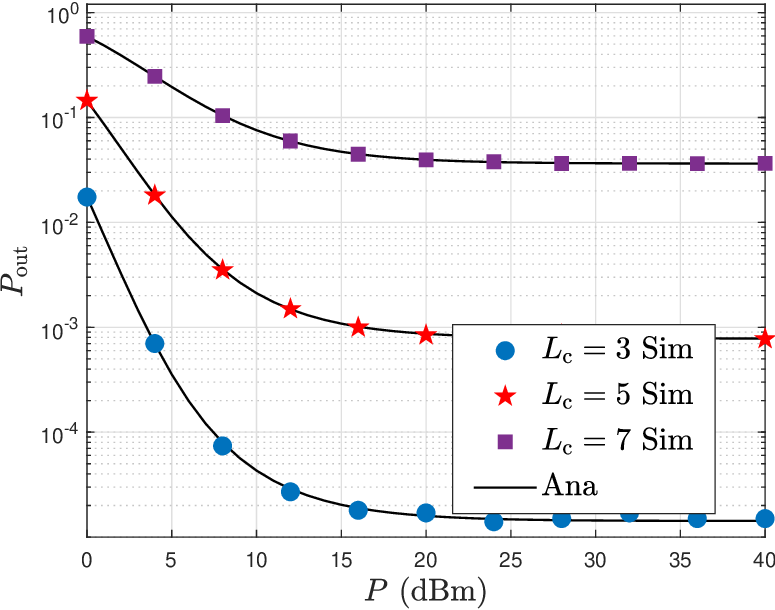}}
	\subfigure[${P_{{\mathrm{sop}},1}^{\mathrm{II}}}$ for varying $L_c$ with ${L_{ec}} = 2$, ${L_{e1}} = L_p$. ]{
		\label{fig3b}
		\includegraphics[width = 0.231 \textwidth]{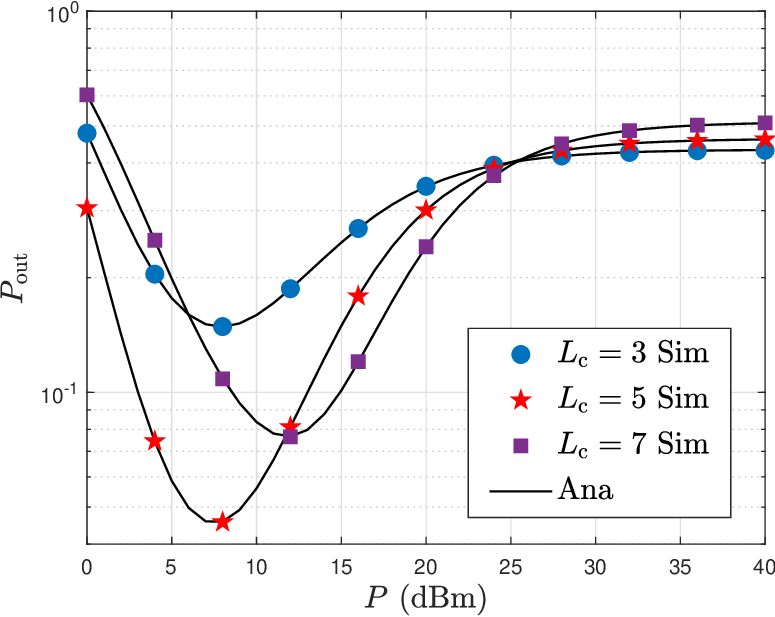}}
	\subfigure[${P_{{\mathrm{sop}},1}^{\mathrm{III}}}$ for varying $L_c$ with ${L_{e1}} = 1$, ${L_{ec}} = L_c$, ${L_{e2}} = L - {L_{ec}} - {L_{e1}}$. ]{
		\label{fig3c}
		\includegraphics[width = 0.231 \textwidth]{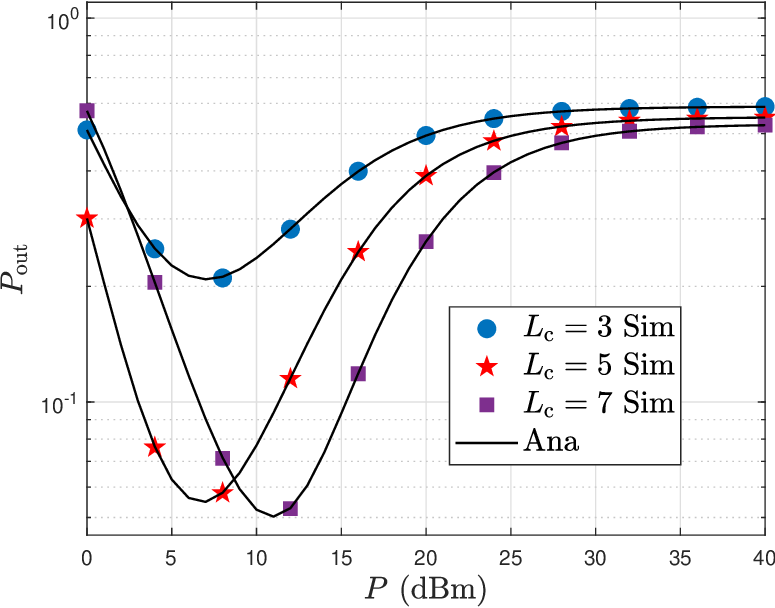}}
	\subfigure[${P_{{\mathrm{sop}},1}^{\mathrm{IV}}}$ for varying $L_c$ with ${L_{ec}} = 2$, ${L_{e2}} = L - {L_{ec}}$. ]{
		\label{fig3d}
		\includegraphics[width = 0.231 \textwidth]{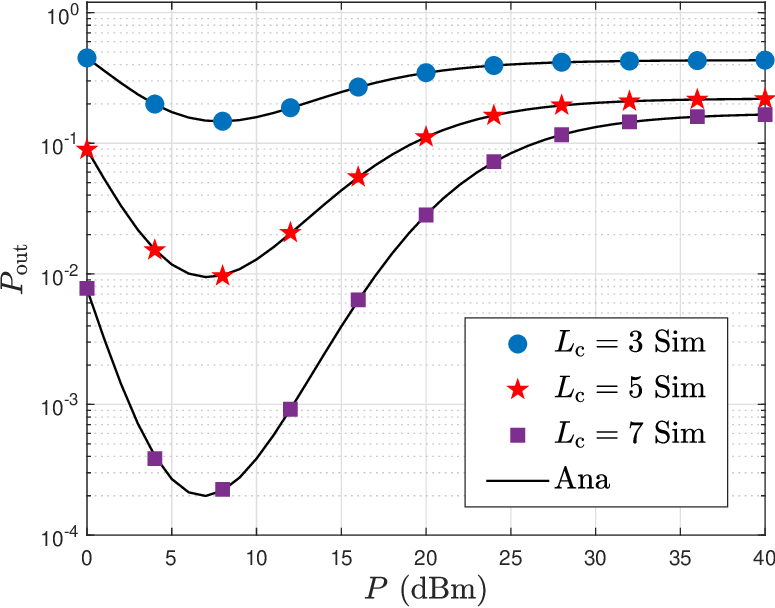}}
	\caption{SOP versus the $P$ with $N_s = 50$, $\tau_c =\tau_1 =\tau_2 = \frac{1}{3}$, $r_1 =15$,  $r_e = 25$, $L = 9$, and $R_{1,c}^{\mathrm{th}} = R_{1,p}^{\mathrm{th}} = 0.1$.}
	\label{fig3}
\end{figure}

Fig. \ref{fig3} presents the impact of $P$ for varying $L_c$ on the SOP of $U_1$.
In Fig. \ref{fig3a}, one can observe the SOP increases as $L_c$ increases, which is easy to follow since more significant $L_c$ leads to less $L_P$, thus, ${\gamma _{1,p}}$ become smaller and transmissions become more vulnerable to intercept.
In Figs. \ref{fig3b} and \ref{fig3c}, one can observe that SOP initially decreases and then increases in the lower-$P$ region as $L_c$ increases, indicating an optimal $L_c$ in the lower-$P$ region to minimize the SOP of $U_1$. 
This is because, in the scenarios with the lower $L_c$, the SOP of $U_1$ depends on both the SINR/SNR of the common and private signals and 
decoding the common stream is the RSMA system's bottleneck; thus, the larger the $L_c$, the larger the ${\gamma _{1,c}}$, and the smaller SOP. 
In the scenarios with the larger $L_c$, decoding the private signals is the RSMA system's bottleneck, and the more significant $L_c$ leads to the smaller ${\gamma _{1,p}}$ and the larger SOP.
In Fig. \ref{fig3d}, it can be observed that the secrecy performance of $U_1$ is improved as $L_c$ increases since larger $L_c$ leads to ${\gamma _{1,c}}$.

\begin{figure}[t]
	\centering
	\subfigure[${P_{{\mathrm{sop}},1}^{\mathrm{I}}}$ for varying $P$ with ${L_{e1}} = 2$. ]{
		\label{fig04a}
		\includegraphics[width = 0.231 \textwidth]{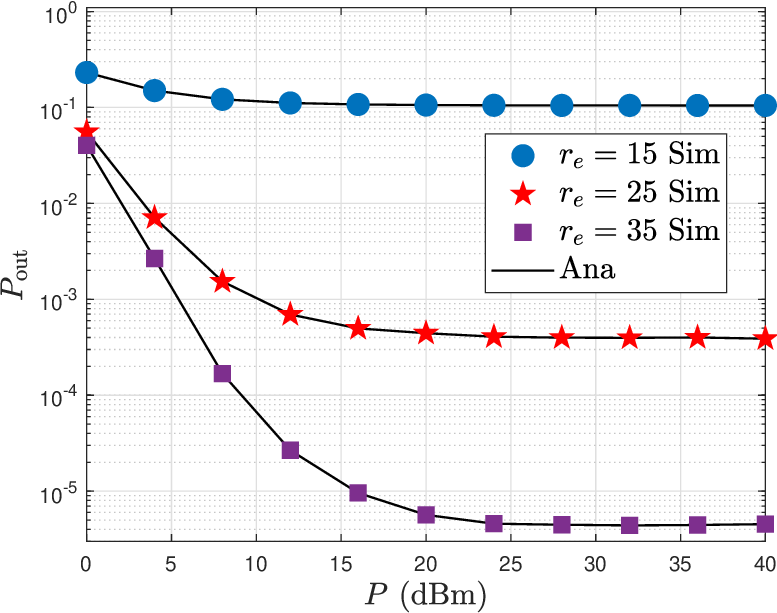}}
	\subfigure[${P_{{\mathrm{sop}},1}^{\mathrm{II}}}$ for varying $P$ with ${L_{ec}} = 2$, ${L_{e1}} = L_p$. ]{
		\label{fig04b}
		\includegraphics[width = 0.231 \textwidth]{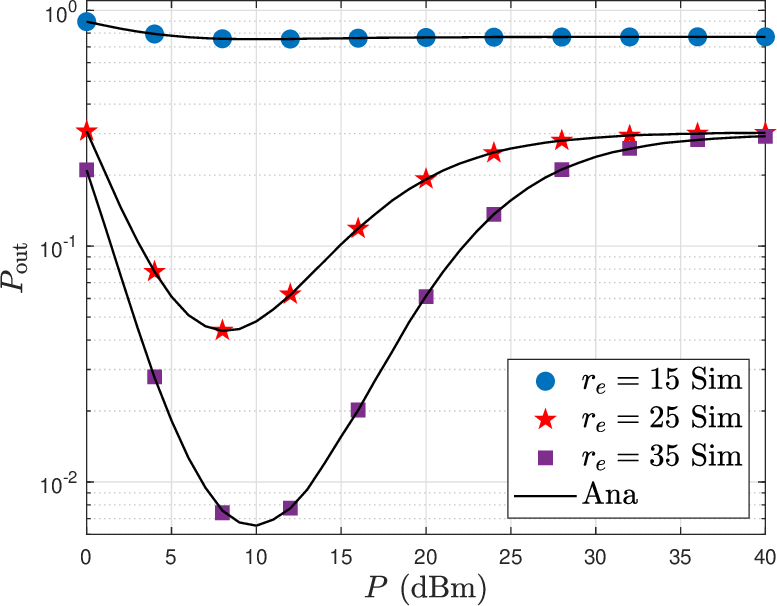}}
	\subfigure[${P_{{\mathrm{sop}},1}^{\mathrm{III}}}$ for varying $P$ with ${L_{e1}} = 1$, ${L_{ec}} = L_c$, ${L_{e2}} = L - {L_{ec}} - {L_{e1}}$. ]{
		\label{fig04c}
		\includegraphics[width = 0.231 \textwidth]{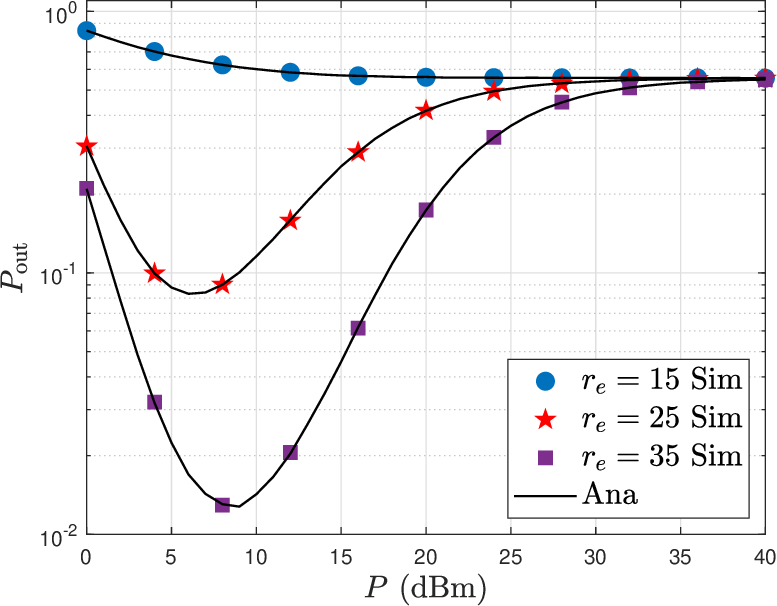}}
	\subfigure[${P_{{\mathrm{sop}},1}^{\mathrm{IV}}}$ for varying $P$ with ${L_{ec}} = 2$, ${L_{e2}} = L - {L_{ec}}$. ]{
		\label{fig04d}
		\includegraphics[width = 0.231 \textwidth]{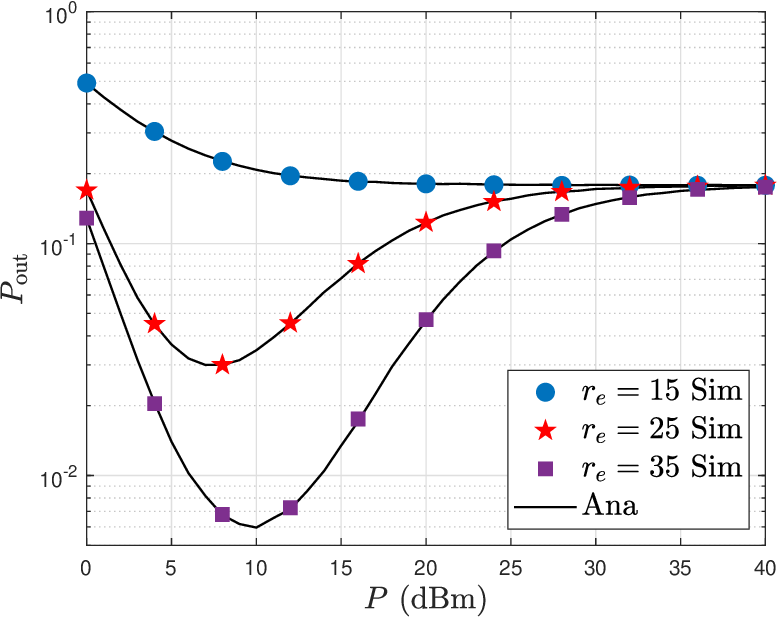}}
	\caption{SOP versus the $\tau_1$ and $\tau_2$ ($\tau_c = 1 - \tau_1 - \tau_2$) with $N_s = 50$,  $r_1 =15$,  $L = 8$, $L_c = 4$, and $R_{1,c}^{\mathrm{th}} = R_{1,p}^{\mathrm{th}} = 0.1$.}
	\label{fig04}
\end{figure}
Fig. \ref{fig04} demonstrates the impact of $P$ with varying $r_e$ on the SOP of $U_1$. {We observe that SOP decreases as $r_e$ increases}; this is because the path loss of $U_1$ is the main factor affecting SOP. 
In the larger-$r_e$ region, SOP decreases as $P$ increases in Fig. \ref{fig04a} while SOP initially decreases and then increases as $P$ increases Figs. \ref{fig04b} - \ref{fig04d} with the same reason as in Figs. \ref{fig2a} - \ref{fig2d}. 
In the low-$r_e$ region, we observe that SOP decreases as $P$ increases in all the cases, which denotes that more power can improve security performance.

\begin{figure}[t]
	\centering
	\subfigure[${P_{{\mathrm{sop}},1}^{\mathrm{I}}}$ for varying $P$ with ${L_{e1}} = 2$. ]{
		\label{fig4a}
		\includegraphics[width = 0.231 \textwidth]{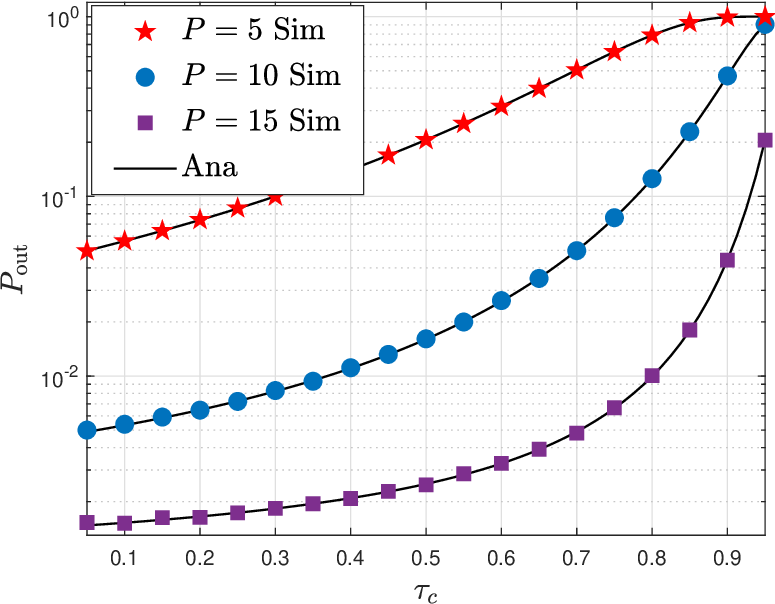}}
	\subfigure[${P_{{\mathrm{sop}},1}^{\mathrm{II}}}$ for varying $P$ with ${L_{ec}} = 2$, ${L_{e1}} = L_p$. ]{
		\label{fig4b}
		\includegraphics[width = 0.231 \textwidth]{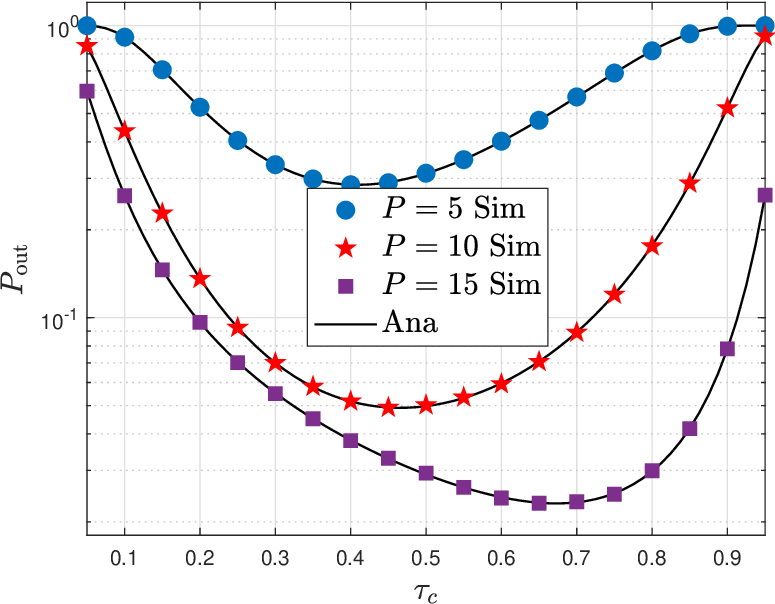}}
	\subfigure[${P_{{\mathrm{sop}},1}^{\mathrm{III}}}$ for varying $P$ with ${L_{e1}} = 1$, ${L_{ec}} = L_c$, ${L_{e2}} = L - {L_{ec}} - {L_{e1}}$. ]{
		\label{fig4c}
		\includegraphics[width = 0.231 \textwidth]{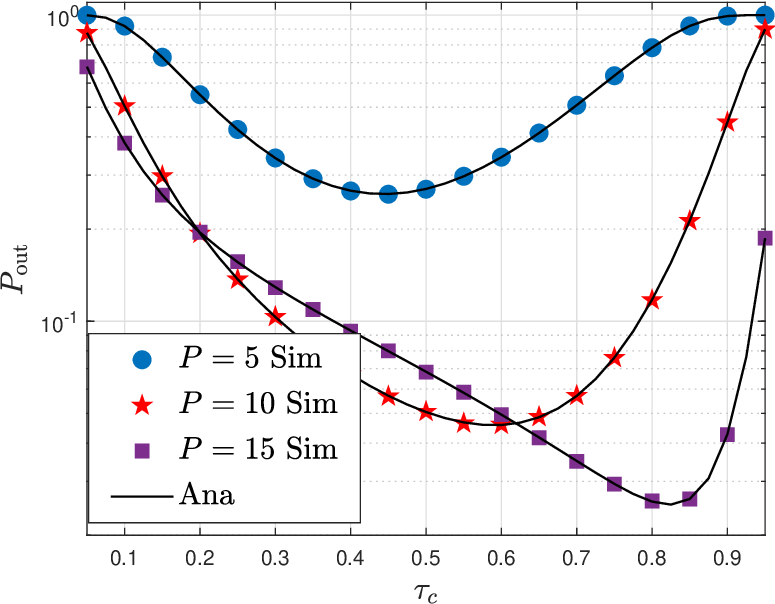}}
	\subfigure[${P_{{\mathrm{sop}},1}^{\mathrm{IV}}}$ for varying $P$ with ${L_{ec}} = 2$, ${L_{e2}} = L - {L_{ec}}$.]{
		\label{fig4d}
		\includegraphics[width = 0.231 \textwidth]{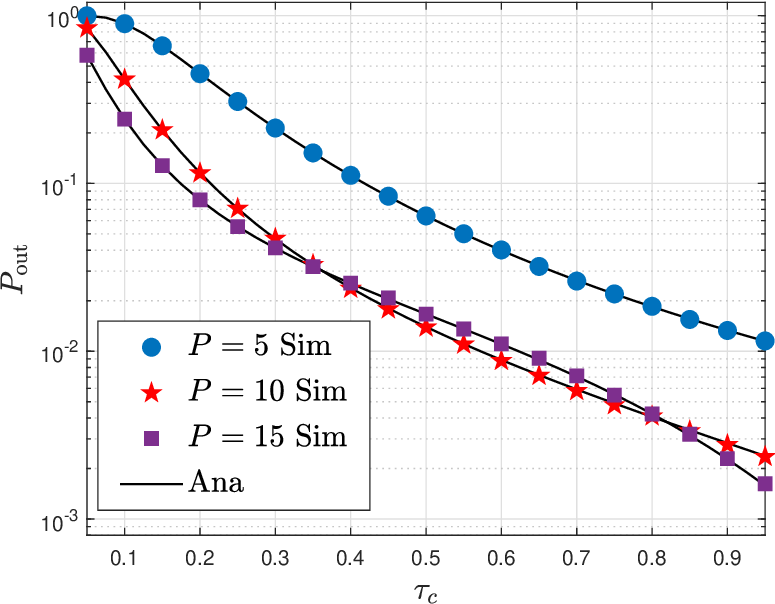}}
	\caption{SOP versus the $\tau_c$ with $N_s = 50$, $\tau_1 =\tau_2$, $r_1 =15$,  $r_e = 25$, $L = 8$, $L_c = 4$, and $R_{1,c}^{\mathrm{th}} = R_{1,p}^{\mathrm{th}} = 0.1$.}
	\label{fig4}
\end{figure}
Fig. \ref{fig4} plots the impact of $\tau_c$ with varying $P$ on the SOP of $U_1$.
We observe that SOP increases as $\tau_c$ in Fig. \ref{fig4a}, which is easy to follow since less power is allocated to private streams as increasing $\tau_c$. 
In Figs. \ref{fig4b} and \ref{fig4c}, one can observe SOP initially decreases and then increases as $\tau_c$ increases, which denotes that there is an optimal $\tau_c$ to minimize the SOP.
This is because in the lower-$\tau_c$ region, decoding the common streams is the bottleneck. 
Thus, increasing $\tau_c$ will enhance the secrecy outage performance.
In the larger-$\tau_c$ region, decoding the private streams will be the bottleneck. 
However, the power allocated to the private stream decreases, so the SOP deteriorates.
Moreover, the optimal $\tau_c$ is relative to $P$ and $L_c$. 
Based on Fig. \ref{fig4d}, the secrecy performance of $U_1$ improves with the increasing $\tau_c$ since the SINR of common streams at $U_1$ increases as $\tau_c$.

\begin{figure}[t]
	\centering
	\subfigure[${P_{{\mathrm{sop}},1}^{\mathrm{II}}}$ for varying $R_{1,p}^{\mathrm{th}}$ with ${L_{ec}} = 2$, ${L_{e1}} = L_p$. ]{
		\label{fig05a}
		\includegraphics[width = 0.231 \textwidth]{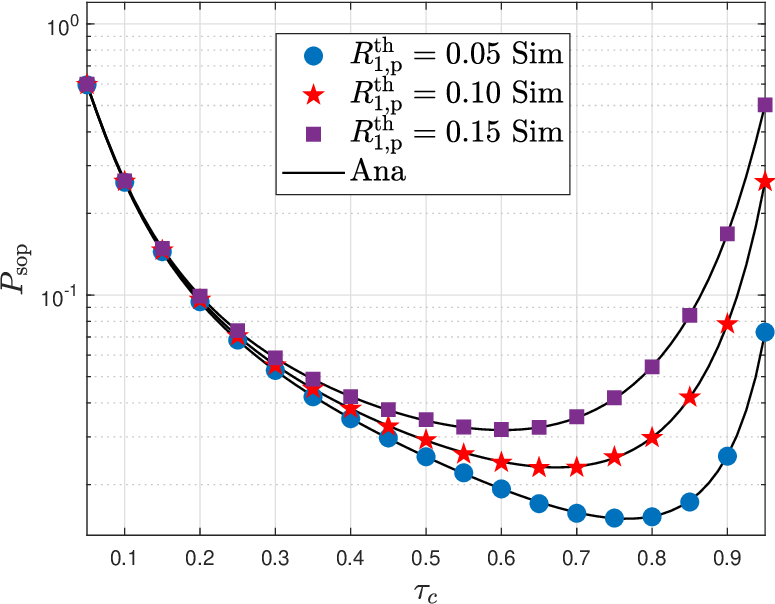}}
	\subfigure[${P_{{\mathrm{sop}},1}^{\mathrm{III}}}$ for varying $R_{1,p}^{\mathrm{th}}$ with ${L_{e1}} = 1$, ${L_{ec}} = L_c$, ${L_{e2}} = L - {L_{ec}} - {L_{e1}}$. ]{
		\label{fig05b}
		\includegraphics[width = 0.231 \textwidth]{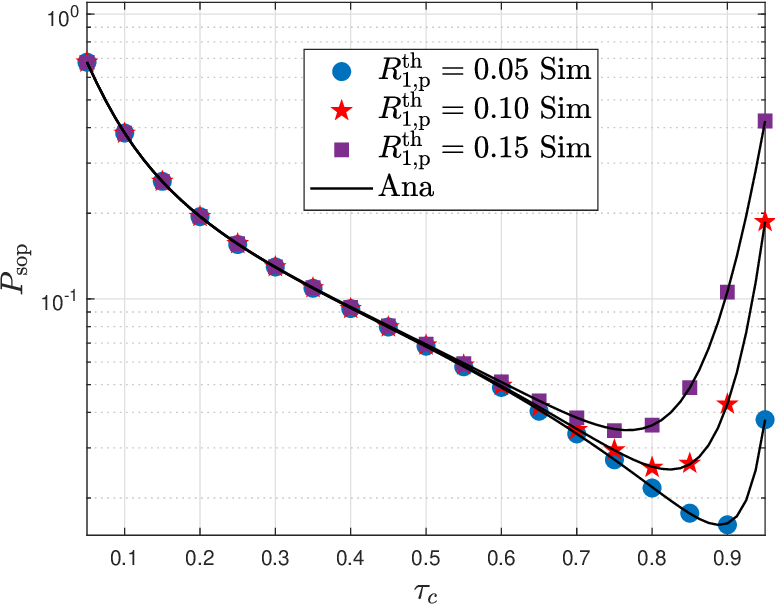}}
	\subfigure[${P_{{\mathrm{sop}},1}^{\mathrm{II}}}$ for varying $R_{1,c}^{\mathrm{th}}$ with ${L_{ec}} = 2$, ${L_{e1}} = L_p$. ]{
		\label{fig05c}
		\includegraphics[width = 0.231 \textwidth]{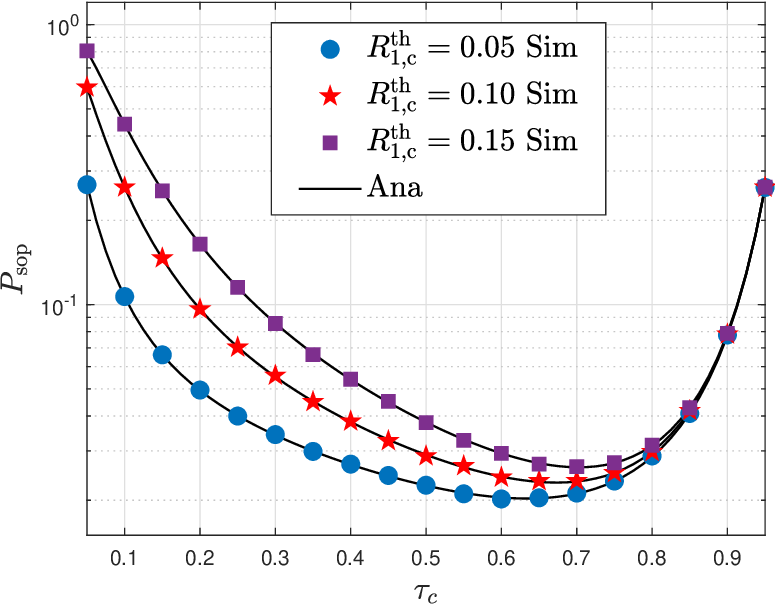}}
	\subfigure[${P_{{\mathrm{sop}},1}^{\mathrm{III}}}$ for varying $R_{1,c}^{\mathrm{th}}$ with ${L_{e1}} = 1$, ${L_{ec}} = L_c$, ${L_{e2}} = L - {L_{ec}} - {L_{e1}}$.]{
		\label{fig05d}
		\includegraphics[width = 0.231 \textwidth]{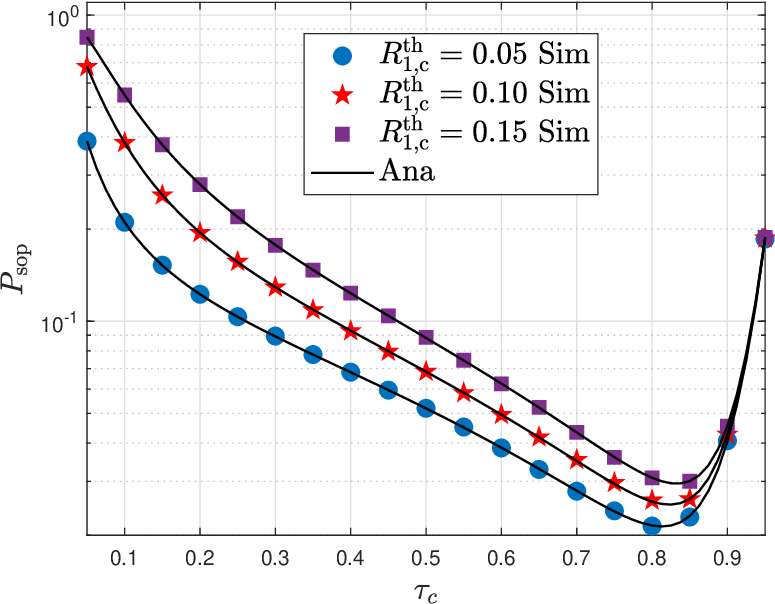}}
	\caption{SOP versus the $\tau_c$ with $N_s = 50$, $\tau_1 =\tau_2$, $r_1 =15$,  $r_e = 25$, $L = 8$, $L_c = 4$, and $R_{1,c}^{\mathrm{th}} = R_{1,p}^{\mathrm{th}} = 0.1$.}
	\label{fig05}
\end{figure}
Fig. \ref{fig05} presents the impact of $\tau_c$ with varying $R_{1,p}^{\mathrm{th}}$/$R_{1,c}^{\mathrm{th}}$ on the SOP of $U_1$.
We observe that SOP decreases as $R_{1,p}^{\mathrm{th}}$/$R_{1,c}^{\mathrm{th}}$ decreases, which is easy to follow since a more significant targeted secrecy rate denotes higher security requirements. 
Moreover, the optimal $\tau_c$ depends on $R_{1,p}^{\mathrm{th}}$/$R_{1,c}^{\mathrm{th}}$. 
The smaller $R_{1,p}^{\mathrm{th}}$/$R_{1,c}^{\mathrm{th}}$, the larger optimal $\tau_c$, which is due to the low requirements and more power can be allocated to the stream to be the system's bottleneck.
In Figs. \ref{fig05a} and \ref{fig05b}, in the large-$\tau_c$ region private signals is the bottleneck, so $R_{1,p}^{\mathrm{th}}$ has a more pronounced impact on the SOP.
However, in Figs. \ref{fig05c} and \ref{fig05d}, in the lower-$\tau_c$ region common signals is the bottleneck, so $R_{1,c}^{\mathrm{th}}$ has a more pronounced impact on SOP.

\begin{figure}[t]
	\centering
	\subfigure[${P_{{\mathrm{sop}},1}^{\mathrm{I}}}$ with ${L_{e1}} = 2$. ]{
		\label{fig7a}
		\includegraphics[width = 0.231 \textwidth]{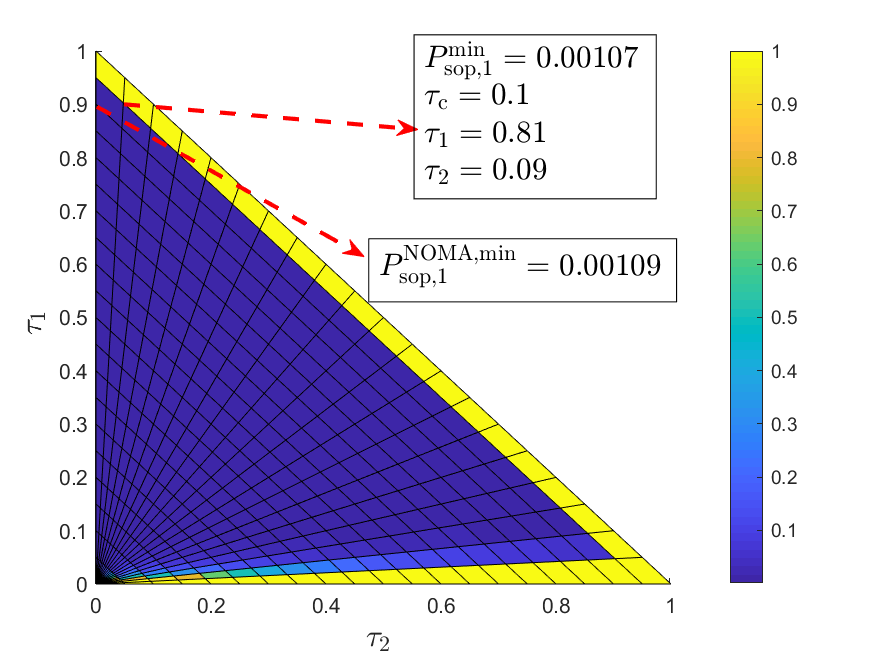}}
	\subfigure[${P_{{\mathrm{sop}},1}^{\mathrm{II}}}$ with ${L_{ec}} = 2$, ${L_{e1}} = L_p$. ]{
		\label{fig7b}
		\includegraphics[width = 0.231 \textwidth]{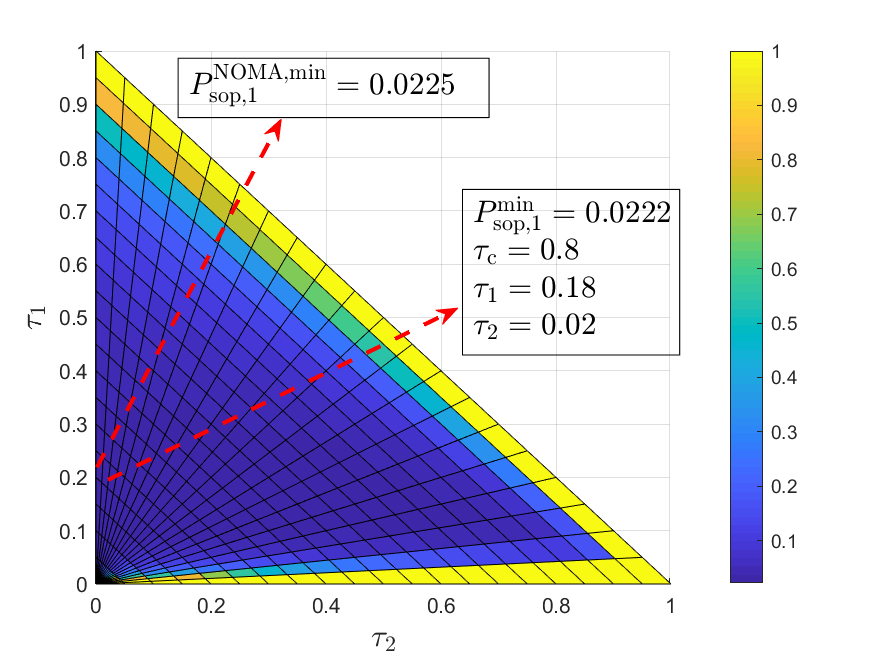}}
	\subfigure[${P_{{\mathrm{sop}},1}^{\mathrm{III}}}$ with ${L_{e1}} = 1$, ${L_{ec}} = L_c$, ${L_{e2}} = L - {L_{ec}} - {L_{e1}}$.  ]{
		\label{fig7c}
		\includegraphics[width = 0.231 \textwidth]{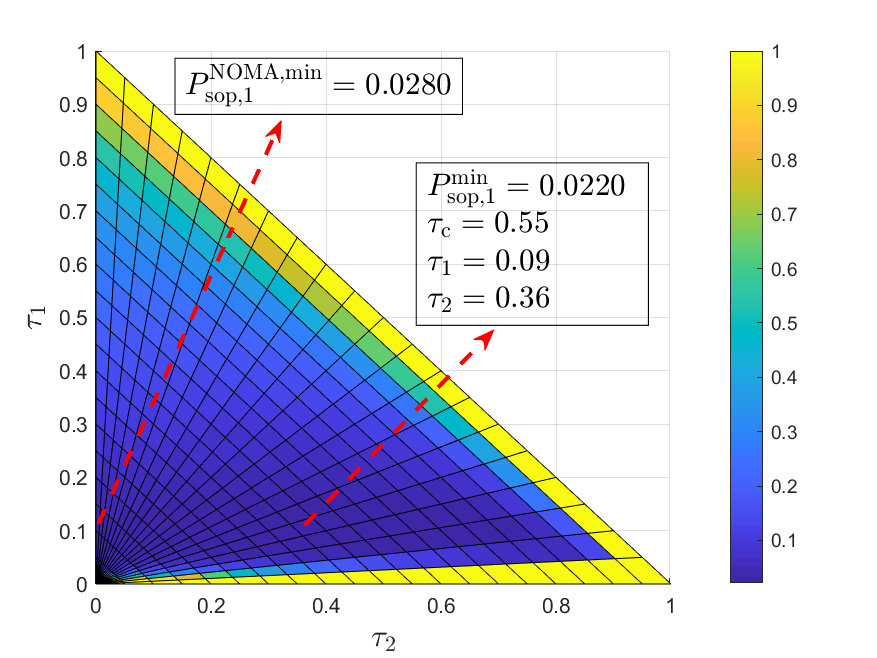}}
	\subfigure[${P_{{\mathrm{sop}},1}^{\mathrm{IV}}}$ with ${L_{ec}} = 2$, ${L_{e2}} = L - {L_{ec}}$.  ]{
		\label{fig7d}
		\includegraphics[width = 0.231 \textwidth]{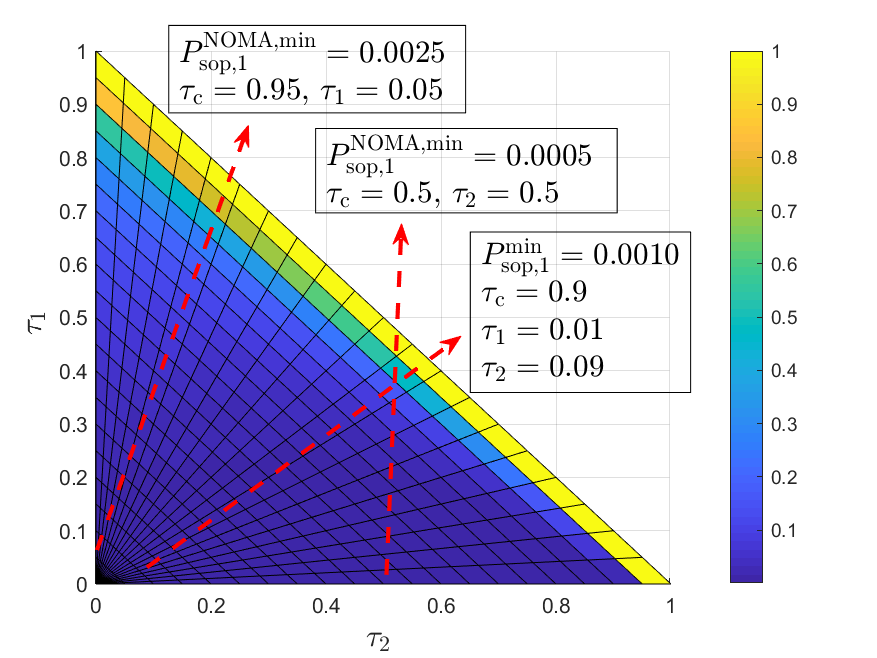}}
	\caption{SOP versus the $\tau_1$ and $\tau_2$ with $N_s = 50$, $P = 10 {\mathrm{th}}{dB}$, $r_1 =15$,  $r_e = 25$, $L = 8$, $L_c = 4$, and $R_{1,c}^{\mathrm{th}} = R_{1,p}^{\mathrm{th}} = 0.1$.}
	\label{fig7}
\end{figure}

{To evaluate the performance of the RSMA-based mmWave systems, NOMA-based mmWave systems is utilized as the benchmark in this work, which can be found in Fig. \ref{fig7}, wherein the SOP of $U_1$ for varying $\tau_c$, $\tau_1$, and $\tau_2$ is presented.
Like \cite{AbolpourM2022OJCS}, $P_{{\mathrm{sop}},1}^{\min }$ and $P_{{\mathrm{sop}},1}^{{\mathrm{NOMA}},{{\min}}}$ represent the least achievable SOP of RSMA and NOMA systems, respectively.}
We observe that $P_{{\mathrm{sop}},1}^{\min }$  successively increases and then decreases from Figs. \ref{fig7a} - Fig. \ref{fig7d}.
The underlying reason is that the overlapped resolvable paths between $U_1$ and $E$ become larger initially, and then eavesdroppers can wiretap more confidential messages.
Then, the overlapped resolvable paths between $U_1$ and $E$ become smaller and also interfered with by messages of $U_2$; thus, the secrecy performance of the mmWave RSMA system is improved.
Fig. \ref{fig7a} indicated that in the case of significant differences of $L_c$ and $L_{e1}$, more power should be allocated to $s_1$ to improve secrecy performance. 
In Figs. \ref{fig7b} - \ref{fig7d}, one can observe that the secrecy performance of common streams is easier to become a bottleneck; thus, larger $\tau_c$ and more minor $\tau_1$ can achieve better secrecy performance.
In Figs. \ref{fig7c} and \ref{fig7d}, a more significant fraction of power should be allocated to $s_2$ {relative to $s_1$} to confuse the eavesdropper.
If $\tau_c$ or $\tau_1$ in Figs. \ref{fig7a} - \ref{fig7c} is close to zero, the SOP would be close to 1 since with a low value of {the users' SINRs} are not sufficient for decoding the common stream or private stream;
When $\tau_2$ equals zero, the RSMA system will degenerate as a NOMA system. We observe that the $P_{{\mathrm{sop}},1}^{\min }$  is less than or equal to $P_{{\mathrm{sop}},1}^{{\mathrm{NOMA}},{{\min}}}$, the underlying reason in Figs. \ref{fig7b} - \ref{fig7c} is that common streams are easier to become a bottleneck. 
More power is allocated to $s_1$ in NOMA systems than RSMA systems, and the secrecy performance of common streams worsens as $\tau_1$ increases.
In Fig. \ref{fig7d}, when $\tau_1$ or $\tau_2$  is equal to zero, the RSMA system would degenerate as a NOMA system, one can observe that $P_{{\mathrm{sop}},1}^{\min }$ at $\tau_2 = 0$ is less than $P_{{\mathrm{sop}},1}^{{\mathrm{NOMA}},{{\min}}}$. In comparison, $P_{{\mathrm{sop}},1}^{\min }$ at $\tau_1 = 0$ is larger than $P_{{\mathrm{sop}},1}^{{\mathrm{NOMA}},{{\min}}}$, the reason is that as $\tau_1 = 0$ {and ${\tau _2} \ne 0$} directly makes ${\gamma _{1,c}}$ increase and ${\gamma _{e,c}^{\mathrm{IV}}}$ decrease.

Based on the results, some new insights are obtained as follows.

\begin{enumerate}		
	\item In RSMA mmWave systems, both common and private streams can be the bottleneck of the security. 
	
	\item There is an optimal transmit power to minimize the SOP of RSMA-based mmWave systems and the optimal transmit power depends on the overlapped paths and distances in the considered systems. 
	
	\item There is an power allocation coefficient to minimize the SOP of RSMA-based mmWave systems and the optimal power allocation coefficient depends on many parameters, such as, the transmit power, target secrecy rates of common streams and private streams, relative location of all the receivers.

	\item Relative to Scenario I and Scenario II, allocating more power to $s_2$ than $s_1$ in Scenario III and Scenario IV can enhance $U_1$'s security.
	
\end{enumerate}

%%%%%%%%%%%%%%%%%%%%%%%%%%%%%%%%%%%%%%%%%%%%%%%%%%%%%%%%%%%%%%%%%%%
\section{Conclusion}
\label{sec:Conclusion}

In this paper, the secure transmissions considering multipath propagation for mmWave RSMA MISO systems were analyzed.
Based on the spatial correlation of the users and the eavesdropper, different eavesdropping scenarios were considered to investigate the
secrecy performance, and then the analytical expressions of the SOP for four scenarios were derived.
Our results illustrated the effects of overlapped paths between receivers, the power allocation coefficient of RSMA users, and channel parameters
on the SOP of RSMA-based mmWave systems. 
{ This work considered that the illegitimate user has the same decode order according to the framework of the RSMA system. 
	Other potential threats or more diverse eavesdropping strategies will be interesting work and part of the future work.}	
Moreover, the results of this work demonstrate an optimal power allocation coefficient to minimize the SOP of RSMA systems. Thus, allocating power reasonably to the common and private streams is vital for the secure transmissions of RSMA mmWave systems.
It is challenging to obtain the analytical expression for the optimal power allocation coefficient to maximize the secrecy performance of RSMA systems. Optimizing the secrecy performance of RSMA mmWave systems will be conducted in our future work.

%%%%%%%%%%%%%%%%%%%%%%%%%%%%%%%%%%%%%%%%%%%%%%%%%%%%%%%%%%%%%%%%%%%

\end{document}